\documentclass[journal]{IEEEtran}

\usepackage{graphicx}
\graphicspath{ {FIG/}}
\usepackage{balance}
\usepackage{enumitem}
\usepackage{amsmath}
\usepackage{amssymb}
\usepackage{listings}
\usepackage{subcaption}
\usepackage{comment}
\usepackage[autostyle]{csquotes}
\usepackage[none]{hyphenat}
\usepackage[dvipsnames]{xcolor}
\usepackage{amsmath}
\usepackage{url}
\usepackage{booktabs}
\usepackage{multirow}
\usepackage{hyperref}
\hypersetup{
    colorlinks=true,
    linkcolor=black,
    filecolor=ForestGreen,      
    urlcolor=blue,
    citecolor = black
}

\begin{document}
%
\title{Defense-in-Depth: A Recipe for Logic Locking to Prevail}

\author{M Tanjidur Rahman$^{1}$, M Sazadur Rahman$^{1}$, 
Huanyu Wang$^{1}$, Shahin Tajik$^{1}$, Waleed Khalil$^{2}$, \\
Farimah Farahmandi$^{1}$, Domenic Forte$^{1}$, Navid Asadizanjani$^{1}$, and Mark Tehranipoor$^{1}$ 
\thanks{$^{1}$Authors are with Florida Institute For Cybersecurity (FICS) Research, Electrical \& Computer Engineering Department, University of Florida, Gainesville, FL, USA. Contact:
{\tt\small mir.rahman$@$ufl.edu, \{nasadi,tehranipoor\}$@$ece.ufl.edu}}\\
\thanks{$^{2}$Author is with Electrical \& Computer Engineering Department, Ohio State University, OH, USA.}}


\maketitle

\begin{abstract}
Logic locking has emerged as a promising solution for protecting the semiconductor intellectual Property (IP) from the untrusted entities in the design and fabrication process. 
Logic locking hides the functionality of the IP by embedding additional key-gates in the circuit.
The correct output of the chip is produced, once the correct key value is available at the input of the key-gates.
The confidentiality of the key is imperative for the security of the locked IP as it stands as the lone barrier against IP infringement.
Therefore, the logic locking is considered as a broken scheme once the key value is exposed. 
The research community has shown the vulnerability of the logic locking techniques against different classes of attacks, such as Oracle-guided and physical attacks.
Although several countermeasures have already been proposed against such attacks, none of them is simultaneously impeccable against Oracle-guided, Oracle-less, and physical attacks.
Under such circumstances, a defense-in-depth approach can be considered as a practical approach in addressing the vulnerabilities of logic locking.
Defense-in-depth is a multilayer defense approach where several independent countermeasures are implemented in the device to provide aggregated protection against different attack vectors. 
Introducing such a multilayer defense model in logic locking is the major contribution of this paper.
With regard to this, we first identify the core components of logic locking schemes, which need to be protected. 
Afterwards, we categorize the vulnerabilities of core components according to potential threats for the locking key in logic locking schemes. 
Furthermore, we propose several defense layers and countermeasures to protect the device from those vulnerabilities.
Finally, we turn our focus to open research questions and conclude with suggestions for future research directions. 



\end{abstract}


\IEEEpeerreviewmaketitle
\section{Introduction}



\IEEEPARstart{O}{ver} the past two decades, the business model for the semiconductor industry has shifted from vertical to horizontal.
In the horizontal model, the original component manufacturers (OCM) outsource different steps of the chip manufacturing process, like intellectual property (IP) design, fabrication, and design-for-test (DFT) structure insertion, to more sophisticated offshore fabrication facilities. 
This approach makes the manufacturing process less expensive for new technology development and scaling down the existing IPs. 
However, due to the number of stakeholders involved in design, integration, manufacturing, and distribution located around the globe, the OCM and IP owner/vendor have lost control over the supply chain. 
As a result, IP piracy, counterfeiting, reverse engineering, and hardware Trojan insertion have become eminent threats in the semiconductor industry.
The conventional passive IP protection methods, e.g., patents and copyrights, provide no protection against the aforementioned threats.
Researchers have proposed several hardware obfuscation techniques, such as logic locking/obfuscation~\cite{EPIC}, state space obfuscation~\cite{harpoon},
and IC camouflaging~\cite{rajendran2013security} as an active approach for safeguarding the IP.

Hardware obfuscation is a method of transforming the design and layout of the IP while maintaining the original functionality of it.
Among the hardware obfuscation techniques, logic locking is emerging as possible solutions for establishing trust in the hardware design.
Logic locking hides the functionality of the chip by inserting additional combinational logic gates~\cite{EPIC} or increasing the state space~\cite{harpoon} in the design.
Logic Locking is a key-based hardware obfuscation approach and the inserted logic elements are generally termed as \textit{key-gates}.
The output of the chip is unlocked once the key-gates are connected to the unlocking key-sequence which configured by the IP owner or OCM through a nonvolatile (NVM) memory after the chip is fabricated.

\begin{figure}[t]
\centering
\includegraphics[width=.9\linewidth]{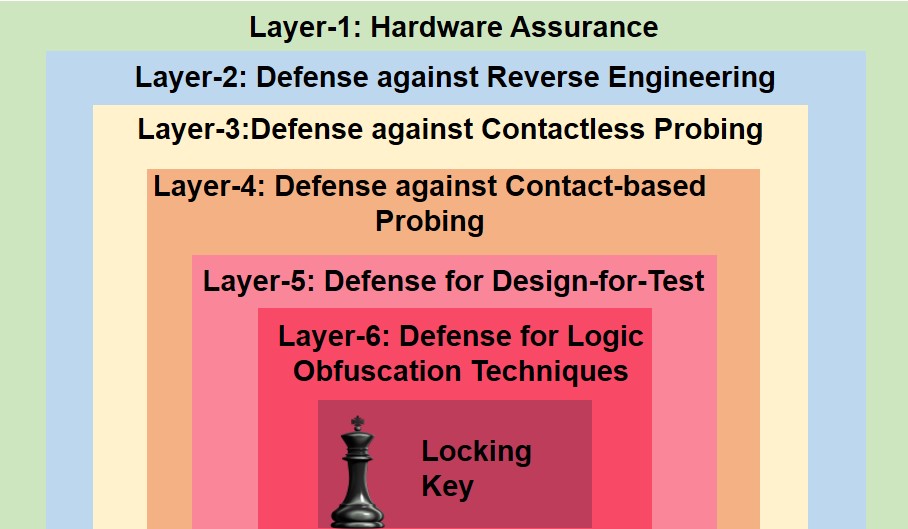} 
\caption{Multiple protection layers in defense-in-depth implementation for logic obfuscation.}
\label{fig:defese_in_depth}
\end{figure}

Although logic locking appeared as a promising protection mechanism against IP piracy, the literature shows that this approach is susceptible to several Oracle-guided attacks, like Boolean Satisfiability (SAT) attacks~\cite{subramanyan2015evaluating, shamsi2017appsat}, Signal Probability Skey (SPS) attacks~\cite{yasin2017removal} and key sensitization attacks~\cite{rajendran2012security}. Over the past several years, the security community has focused on developing countermeasures to hinder those Oracle-guided attacks~\cite{xie2018anti, yasin2016sarlock}. 
Although protection against the above-mentioned attacks received a lot of attention, unfortunately, the security of the key itself is still ignored.
The reason for such ignorance is lying under the two common assumption made in those aforementioned attacks.
First, as the untrusted foundry does not possess the key during fabrication and has only access to the locked netlist/layout and the scan chain, implemented as design-for-test (DFT), only Oracle-guided attacks are considered as the most acceptable method of key extraction.
Second, the unlocking key is written into a \textit{tamper-} and \textit{read-proof} memory, and therefore, is protected against reverse engineering in the field. 
However, an adversary such as an untrusted foundry with access to most advanced failure analysis (FA) equipment, such as microprobing station, scanning electron microscope (SEM) and laser scanning microscope (LSM), should be more than capable of extracting the unlocking key from a chip by contact-based electrical~\cite{wang2016probing, helfmeier2013breaking} or contactless optical probing~\cite{tajik2017power}.
Furthermore, the literature on logic locking does not consider the threat imposed by an end user with full-blown reverse engineering capability~\cite{cryptoeprint:2019:719}.
The task of a reverse engineer can be made difficult through implementing physical layout obfuscation techniques like camouflage cells, dummy vias, filler cells, etc. in the chip~\cite{cocchi2014circuit, rajendran2013security}. 
However, the aforementioned layout obfuscation methods do not eliminate the threat of IP piracy by reverse engineering.
Fig.~\ref{fig:defese_in_depth} shows all the possible attacks against logic locking as well as required defense mechanisms to address them.
Thus, key-based obfuscation techniques are less secure against physical attacks than previously thought due to the possibility/ease of key extraction.
As a result, after nearly a decade of research, none of the logic locking techniques are able to provide impeccable defense against IP piracy/theft and root-of-trust violation.

The security measures developed for IP protection have always been a one-to-one exercise, where a security designer deploys specific technology to counter a specific risk or attack.
However, \enquote{hackers} are innovative and can bypass any security measure implemented in the chip. Therefore, developing a layered defense approach, known as \textit{defense-in-depth}, can be a more practical approach for addressing the security challenges in the hardware security domain.
The similar idea has also been implemented in the cybersecurity community to detect and prevent malicious intruders in a system.
A multilayered defense-in-depth approach, as shown in Fig.~\ref{fig:defese_in_depth}, developed for a logic locked device, can defend the locking key value in an obscured system against any attack by deploying several independent protection layers and eventually raising the cost of all attacks to unacceptable levels.
Multiple defense layers also reduce the probability of intrusion through any other backdoor which was left open unintentionally.




\paragraph*{Contribution}
The main goal of this paper is to introduce a multi-layer protection approach (defense-in-depth) for various threats against logic locking. 
Developing a multilayered defense for logic locking requires an analysis of the vulnerabilities in an obfuscated chip.
Understanding of such security weaknesses contributes to developing a comprehensive threat model; attackers intent, capability, and opportunity analysis.
Depending on the threat model, an OCM can select the appropriate layers for implementing the defense-in-depth for the device.
Therefore, we first identify the core components in logic locking schemes, and explain the idea of defense-in-depth. 
The design steps for developing a multi-layer defense to address the existing vulnerabilities of the logic obfuscation is also explained. 
Then, we describe the vulnerabilities of the core components in the locked chip.
A comprehensive analysis of susceptibilities at different stages of the supply chain is presented as well.
Such analysis facilitates the developing of threat models for different adversaries.
Based on the vulnerability analysis and threat model, we propose a six-layer security architecture for developing the defense-in-depth concept.
Consequently, an in-depth survey of the existing security countermeasures, best practices, and standards depending on the assets defending at each defense layer is presented.
Finally, a framework for developing a multi-layered defense-in-depth for hardware obfuscation is outlined for future work.

Defense-in-depth for hardware obfuscation can be commonly compared with the \enquote{castle approach} as it mirrors the layered defenses in a medieval castle to protect the \enquote{king} from an attacker. 
In an obfuscated hardware, the unlocking key is considered as the king in the chip.
Hence, the functionality of the chip is protected by holistic and multiple layered defense scheme implemented as defense-in-depth (Fig.~\ref{fig:defese_in_depth}).

The paper is organized as follows. In Sect.~\ref{Sec:background}, we discuss the basics of hardware obfuscation and logic locking. In Sect.~\ref{Sec:corecomponent} and~\ref{sec:defense_in_depth} the core components in a locked device are identified and the idea of defense-in-depth is introduced, respectively. 
We presented the the susceptibilities of the core components in Sect.~\ref{sec:defense_in_depth}. 
Afterward, in Sect.~\ref{sec:supply_chain}, we explore the existing vulnerabilities of the IC manufacturing process and supply chain and explain threat models for different potential adversaries.
The architecture of the defense-in-depth model for the obfuscated chip is presented in Sect.~\ref{sec:model}.
The available countermeasures to thwart the threat against the existing attacks at different layers of defense and security standards are reviewed in Sect.~\ref{sec:countermeasure}.
The future research opportunities for developing the security of hardware obfuscation are discussed in Sect.~\ref{sec:future_research}. Finally, we conclude the paper in Sect.~\ref{sec:conclusion}.

\section{Background} \label{Sec:background}

\subsection{Hardware Obfuscation}
The objective of the hardware obfuscation is twofold -- a) concealing the design secret, such as the algorithm and implementation, against reverse engineering and b) making the design unusable as a blackbox and unintelligible for IP piracy.
This obscurity can be achieved through changing certain nodes, embedding additional logic gates, altering state-transition-graph or manipulating device or interconnect layers~\cite{ cocchi2014circuit, rajendran2013security, EPIC,  harpoon}. 
Obfuscation methods can be classified into three categories based on the design stage at which the obfuscation is performed~\cite{amir2017comparative}.


\subsubsection{Pre-synthesis Obfuscation} Pre-synthesis obfuscation is applied on register-transfer-level (RTL) IPs, which are commonly known as soft IPs.
A Soft IP is usually offered in a high-level language like C++, Verilog, or VHDL form.
In the case of pre-synthesis obfuscation, the IP is encrypted with well-known encryption techniques, e.g., IEEE P1735~\cite{ieee1735}.
Obfuscating the RTL code with a finite state machine (FSM) has also been proposed, where the code later traversed with a key sequence or code-word~\cite{amir2017comparative}.
\subsubsection{Post-synthesis Obfuscation} post-synthesis obfuscation is the method of hiding the true functionality of the device under attack (DUA) through the insertion of additional logic elements or modification in the FSM.
The primary objective of structural obfuscation is preventing IP piracy.
Combinational logic locking and FSM locking are two most researched post-synthesis obfuscation methods in the literature.

\subsubsection{Physical Layout Obfuscation} 
\begin{figure}[t]
\centering
\includegraphics[width=.9\linewidth]{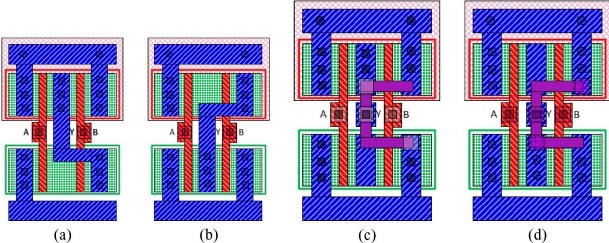} 
\caption{Standard NAND gate (a) and NOR gate (b). These gates could be easily differentiable by looking at the top metal layers. Camouflaged NAND gate (c) and NOR gate (d). These gates have identical top metal layers and are therefore, harder to identify \cite{rajendran2013security}.}
\label{fig:camouflage_cells}
\end{figure}

The objective of physical layout obfuscation or layout obfuscation is to thwart the IP reverse engineering and prevent any malicious modifications in the layout.
In this method, the physical characteristics of the circuit or the layout is modified to increase ambiguity in cell identification or connectivity. 
Several techniques have been proposed for layout obfuscation, such as doping based techniques, and dummy contact insertion in the fabrication level ~\cite{becker2014stealthy}.
The layout can also be hidden at the cell level using camouflaging cells~\cite{rajendran2013security}.
Camouflage cells alter the layout of two standard cells with different functionalities to appear identical.
Camouflage cells can be developed using real and dummy contacts.
As shown in Fig.~\ref{fig:camouflage_cells}(a) and \ref{fig:camouflage_cells}(b), 2-input NAND and NOR gates can be differentiated through analyzing the active region and metal layers.
These two gates can be made looked identical (Fig.~\ref{fig:camouflage_cells}(c) and \ref{fig:camouflage_cells}(d)) by introducing dummy vias.
Inserting dummy gates, dummy filler metal or manipulating doping implant have also been used to generate camouflage cells~\cite{cocchi2014circuit, shamsi2017cyclic, shakya2019covert}.
The insertion of dummy vias and identical logic gates introduce ambiguity in image processing based reverse engineering.
However, camouflage cells introduce area, power, and delay overhead in the design~\cite{yasin2016camoperturb}. 
In the case of gate-level obfuscation, camouflage cell insertion algorithms~\cite{rajendran2013security} have been proposed.
Camouflage connections~\cite{shamsi2017cyclic} and vanishing vias~\cite{bhunia2018vanishing} have also been proposed to prevent reverse engineering.

\begin{figure}[t]
\centering
\includegraphics[width=1\linewidth]{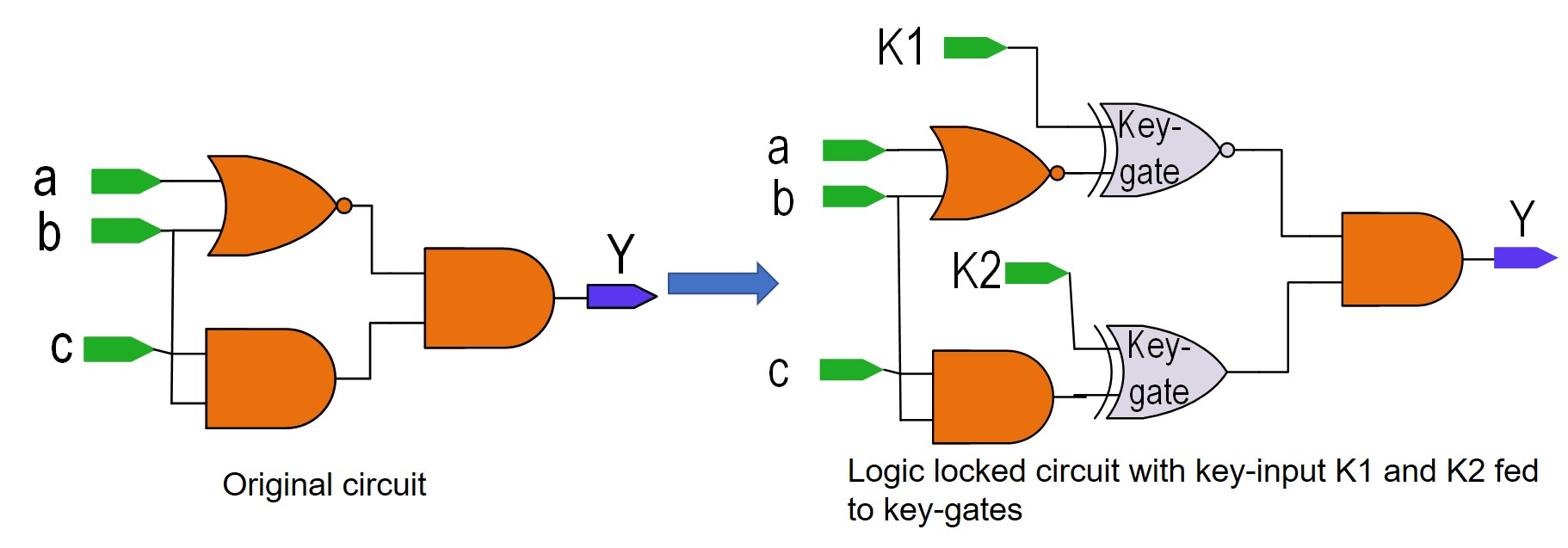} 
\caption{Simplified example of logic locking method.}
\label{fig:logic_locking_example}
\end{figure}



\subsection{Logic Locking} Logic locking or logic obfuscation is developed to hide the functionality of an IP by inserting additional logic gates into the netlist of IP.
Such protection is provided through embedding additional logic gates into the combinational or sequential parts of the design (Fig.~\ref{fig:logic_locking_example}).
While the former approach is called combinational logic locking, the latter is called FSM locking.
In the case of combinational logic locking, the extra embedded logic gates are known as \textit{key-gates}, which are connected to primary inputs that are collectively referred to as the \textit{key}.
On the other hand, in FSM locking approaches, the functionality of the IP is obscured with additional states in the state transition graph~\cite{harpoon}.
Applying a correct sequence of the key, an authorized user can initiate the functional state of the IP/chip. 
In both techniques, the design provides the correct functionality only if the provided key-input values are correct.
Otherwise, the IP does not reveal correct input-output behaviour.
The key value is only available to the OCM and the IP owner and not available during the fabrication process.
Therefore, once the chips are fabricated, they are transferred to a trusted facility for programming the key, known by the design house, into a secure and tamper-proof \textit{key-storage element}.
In the case of combinational logic locking, it has already been shown that random insertion of key-gates may not add a significant security feature to the design~\cite{rajendran2015fault}.
Therefore, several key-gate insertion algorithms, like the insertion of XOR/XNOR gates~\cite{EPIC, rajendran2015fault}, lookup tables~\cite{baumgarten2010preventing}, and multiplexers~\cite{rajendran2015fault} have been proposed. 

\section{Core Components in an Obfuscated IC} \label{Sec:corecomponent}
In this section, we discuss the core components of a locked device.
Each component is defined by its functions and involvement in the security of the device.
An IC implemented with either combinational or sequential logic locking have five imperative components -- (a) Key-storage element; (b) Key-delivery unit; (c) Interconnects; (d) Design-for-test; (e) Obfuscated hardware.

\begin{figure}[t]
\centering
\includegraphics[width=1\linewidth]{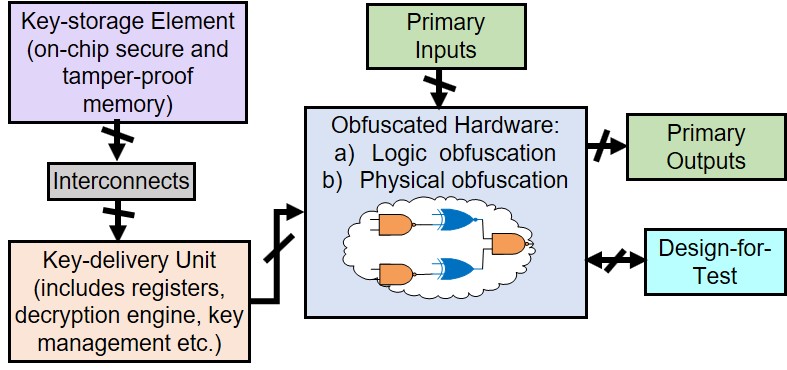} 
\caption{Core components in an IC implemented with hardware obfuscation. }
\label{fig:logic_locking_block}
\end{figure}

\subsection{Key-storage Element} \label{sec:core_key_storage}
In logic locking, the key value is not available during the fabrication process. 
After the fabrication, the ICs are transferred to a trusted facility for configuring the key into a secure and tamper-proof \textit{key-storage element} (see Fig.~\ref{fig:logic_locking_block}). 
As the key is essential for the correct functionality of the device, storing the key in volatile memory is not suitable for such a purpose. 
In the case of a volatile key-storage, keeping the chip in a continuous power-up state to maintain the stored value is not a practical approach in terms of power consumption. 
Therefore, non-volatile memories (NVMs) are the conventional choice as key storage elements. 
Flash, EEPROM, eFuse, antifuse, and BBRAM are examples of such NVMs.


\subsection{Interconnects}
\textit{Interconnects} are the metal wires in the chip which connect different elements, like transistors, capacitors, etc. and naturally more complex modules, such as memory, processors, cache, etc. in the chip. 
Depending on the functionality and complexity of the IC, the number of interconnect layers may vary.
All devices exchange confidential data between memory and other operational units in the chip through interconnects. 
For example, The obfuscation keys and other security-critical assets, such as encryption keys, device configuration, and manufacturer firmware are typically stored in a key-storage memory cells.
Therefore, these memory cells storing the assets are the root of the security for the design, which needs exclusive protections, such as memory encryption techniques. 
However, to process the assets in the logic, they have to be transmitted to the logic parts of the chip through chip interconnects.
Hence, protecting the interconnects against potential vulnerabilities, such as probing and bus snooping, is equally important in logic obfuscation schemes.

\subsection{Key-delivery Unit}
The key value is compulsory for the operation of the corresponding key-based obfuscated IP. 
Hence, initialization of any IP must include reading the locking key from the key-storage element. 
Thereafter, the key must be fed to the key-gates through registers connected to those key-gates~\cite{cryptoeprint:2019:719}. 
These registers, which can be termed as key-registers, should be privileged registers to prevent any inadvertent manipulation of key values and should maintain the stored data during the entire operating period of the IP/chip.
Moreover, the unlocking key can also be stored in an encrypted format in the key-storage \cite{EPIC}. 
The encrypted key must be decrypted before fed to the key-gates.
This implies the involvement of a decryption engine.
Furthermore, reading the key from secured storage may include key-management logic in the chip for cryptomodules. 
All the key-read circuitry, key-registers, and key-management logic establish the \textit{key delivery} unit for a locked device and should be protected against asset leakage.

\subsection{Design-for-Test}
\textit{Design-for-Test} techniques are widely used in modern system-on-chips (SoCs) to ensure testability of internal circuit elements for monitoring the reliability of the hardware design. This added feature makes it easier to perform structural tests in the hardware design. The manufacturing process is not perfect, making post-silicon validation of designed hardware a vital one. The purpose of functional tests is to verify the correct functionality of the hardware design. However, functional tests are very expensive and the complexity of applying them is too high to realize. To circumvent this obstacle, additional DFT logic is added in the circuit to overcome the difficulty of functional testing in a divide and conquer fashion. For all these obvious reasons, we are considering DFT as a core component in an obfuscated IC.
Design-for-test can be inserted in the design by replacing sequential memory elements with scan cells and converting a sequential design into a combinational one to facilitate the structural testing process. 
However, these scan cells can be used to attack obfuscated hardware designs to extract keys, e.g., key sensitization and Oracle-based attacks. 



\subsection{Obfuscated Hardware}\label{sec:core_component_obfuscated_hardware}
The last core element for the security of the chip is the \textit{obfuscated hardware}.
The functionality and layout of the chip can be concealed from an adversary by implementing different logic locking and physical obfuscation techniques.
Depending on the objective of the hardware obfuscation, the obfuscation techniques can be applied in three ways:
\subsubsection{Device-level Hardware Obfuscation}
At the device level, the layout of the device is disguised by introducing stuck-at-fault or delay manipulation~\cite{vijayakumar2017physical}.
Changes in doping concentration, manipulating inter-layer dielectric, inserting dummy logic and interconnects are conventional techniques to achieve a device-level obfuscated hardware. 
\subsubsection{Circuit-level Hardware Obfuscation} 
The circuit-level hardware obfuscation focuses on hiding the gate functionality by modifying cell libraries~\cite{vijayakumar2017physical}.
Camouflage cells, filler cell, dummy vias, and dummy interconnects are examples of circuit-level obfuscation.  
\subsubsection{System-level or Gate-level Hardware Technique}
Logic locking techniques, i.e., combinational logic locking and FSM locking are considered as system-level or gate-level obfuscation techniques.
The algorithms used for structural and physical obfuscation methods are also considered as system-level techniques for obfuscating the chip design.


\section{Defense-in-depth} \label{sec:defense_in_depth}
 \begin{figure*}[htb] 
\centering
\includegraphics[width=0.75\textwidth]{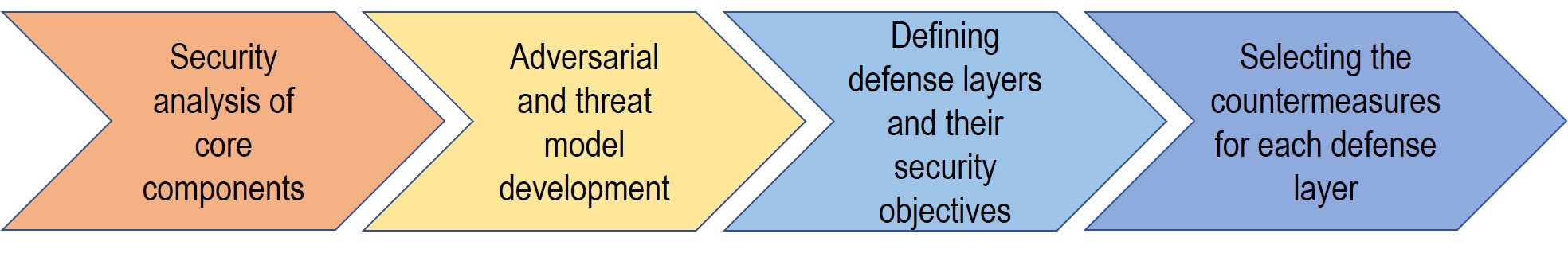}
\caption{Steps for developing a defense-in-depth model for logic locking.}
\label{fig:planning}
\end{figure*}
\subsection{Motivation and Definition of Defense-in-depth}
 The vulnerabilities of core components leave a wide attack surface available for different adversary to extract the assets, i.e., the locking key, layout, and design implementation, from the IC.
 Naturally, a single defensive mechanism against a specific vulnerability cannot protect the functionality and design of the chip against all potential threats.
Once an attacker bypasses the only defensive mechanism implemented in the chip, the security of the entire locking mechanism is broken.
For instance, developing mitigation against oracle-guided attacks, namely SAT attacks, cannot defend against the threat of physical attacks, like optical and electrical probing.
As a result, multiple layers of countermeasures should be implemented to provide protection for the IP/chip against a wide range of attack vectors.
Such a multi-layer defense approach is identified as defense-in-depth.
In this paper, we present the defense-in-depth model where different layers of security system address different vulnerabilities of core components. 

\subsection{Developing the Model for Defense-in-depth}
Developing a model for in-depth defense mechanism for logic obfuscation requires a complex set of analysis on interconnections and dependencies between the different aspects of the supply chain, threat model, system design, the protection mechanism, and assets. 
Besides, providing effective monitoring and protection is required for mitigating the attacks on the IC. 
Developing a defense-in-depth model for hardware obfuscation can be compiled in four stages as shown in Fig.~\ref{fig:planning};

\paragraph{Security Analysis of Core Components}
The first step for modeling the defense-in-depth is identifying the vulnerabilities that are present in the core components of logic locking. 
The assets and methodology of extracting key and design implementation form an obscured chip, i.e., the attack surface of the IC is identified at this stage. 
\paragraph{Threat Model Analysis} 
In developing countermeasures and standards for protecting IPs from piracy, overbuilding, or hardware Trojan insertion, the capability of the adversary has been critically underestimated. 
An attacker can exploit any existing vulnerability in the design which may remain undetected for a long period of time. 
Therefore, assessing the roles of the stakeholders in the supply chain facilitates in identifying the presence of potential adversaries in the supply chain. 
The attack surface can also be defined using the vulnerability analysis of supply chain. 
Analyzing the capabilities, goals of an adversary, and availability of assets is another dimension for selecting the attack mythology and significantly influence the defense-in-depth modeling. 
\paragraph{Developing the Defense-in-Depth Architecture} At this stage, the designer defines the defense layers that protect the chip assets (for example, the defense-in-depth layers depicted in Fig.~\ref{fig:defese_in_depth}) based on the vulnerabilities of core components, the threat model, desired level of security and design budget, e.g., the cost of area, power, and energy for the security of the design secrets. 
A malicious entity can gain unauthorized access to design assets through the simple shortcomings in the design architecture perimeter, or embedded capabilities in the design that are forgotten, unnoticed, or simple disregarded.
Therefore, a multi-layer defense apporach must address the protection for the aforementioned 'backdoors' in the device. 

\paragraph{Security Standards and Selection of Countermeasures} The next step for developing defense-in-depth is to identify the effective countermeasures and protection schemes for protecting core components from the adversary. 
Design budget, i.e., area, power, and energy, defined at the architecture stage plays definitive role in the selection of countermeasures.

\section{Security Analysis of Core Components} \label{sec:vulnerability}
Although most research efforts have been confined to protect the obfuscated IC by improving the security of obfuscated hardware and DFT, a comprehensive study about the possible vulnerabilities of other core elements in hardware obfuscation is still absent in the literature. 
In this section, we will discuss the vulnerabilities of the core elements in an obfuscated IC.

\subsection{Vulnerabilities of the Key-storage Element}
\label{sec:key_storage_vul}
\begin{figure*}[t]
        \centering

        \begin{subfigure}[b]{0.2\textwidth}
                \includegraphics [width=\textwidth]{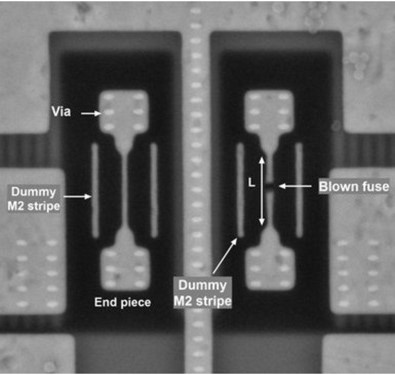}
                \caption{}
                \label{fig:eFuse}
        \end{subfigure}%
        \hspace{3pt}
        \begin{subfigure}[b]{0.35\textwidth}
                \includegraphics [width=\textwidth]{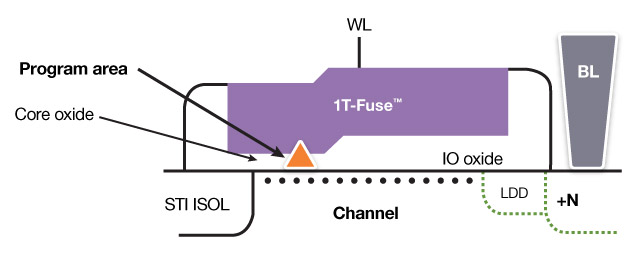}
                \caption{}
                \label{fig:antifuse}
        \end{subfigure}
        \hspace{1pt}
         \begin{subfigure}[b]{0.35\textwidth}
                \includegraphics [width=\textwidth]{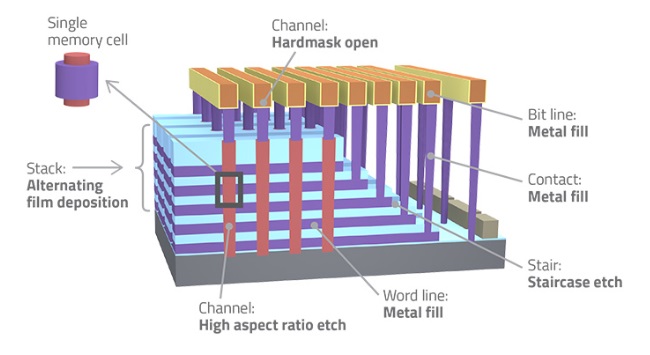}
                \caption{}
                \label{fig:3D_NAND}
        \end{subfigure}
\caption{(a) Difference between before and after program of a TSMC eFuse structure in Qualcomm Gobi MDM9235 Modem 20 nm HKMG \cite{semiengineering_1}; (b) 1T-Fuse Bit Cell in DesignWare OTP NVM IP. The cell is programmed by applying a  controlled, irreversible breakdown voltage from the gate through the core (gate) oxide to the channel \cite{synopsys}; (c) Key process steps for 3D Nand fabrication process \cite{3dnand}. }
\label{fig:key_storage}
\end{figure*}

Protecting the key-storage element is vital for logic locking schemes since the exposure of unlocking key breaks the security of the entire scheme. 
NVMs, like ROM, EEPROM, and Flash, are the prominent candidates for key-storage. 
The NVM can be realized as off-chip or on-chip memory. 
As off-chip memory is vulnerable to data interception attack at chip boundary, on-chip NVM is the only suitable choice as secure key storage. 
Although aforementioned memory technologies are widely deployed by the industry as secure and tamper-proof memories, the main vulnerability of NVM is the availability of the data stored in the memory during the power-off state.
In this state, the memory remains defenseless against any tampering attack.
Therefore, an adversary can deploy advanced FA tools to reverse engineer the memory and readout its contents.

One option for securing key-storage is One Time Programmable (OTP) memory, such as ROM, electric fuse (eFuse) and antifuse.
OTP facilitates to configure the device before shipping to the end user once the chip is fabricated.
eFuse is a continuous metal or polysilicon shape etched on the silicon surface. 
An eFuse structure is shown in Fig.~\ref{fig:eFuse}. 
When a voltage is applied to the eFuse, electromigration causes the open circuit in the cell (the broken fuse in Fig.~\ref{fig:eFuse}) and program the eFuse~\cite{semiengineering_1}.
An attacker with access to FA tools can deprocess the entire die and locate the location of eFuse. 
Later, using the scanning electron microscopy (SEM), she can differentiate between the programmed and unprogrammed eFuse link by observing the metal or silicide link of the eFuse.   
Similar information can be extracted using electrical probing. 
On the other hand, due to scalability into 7~nm node technology, relatively smaller antifuse cells appear as promising solutions to key-storage element.
Antifuse is a standard CMOS transistor which acts as a high resistance in its unprogrammed state. 
Once electrical stress is applied to the gate oxide of the transistor (see Fig.~\ref{fig:antifuse}), the transistor acts as a low resistance conductive path. 
Antifuse can also be placed as via between two metal lines in the chip.
In such a case, detecting the location of antifuse is difficult with SEM imaging.
SEM provides information about the die surface, i.e., the XY plane of the die. 
However, the lateral information of the metal layers in the die is required to distinguish the antifuse fabricated as via. 
The lateral information of the metal layers can only be observed by transmission electron microscopy (TEM). 
As sample preparation and imaging for TEM are more challenging than SEM, differentiating between the programmed and unprogrammed bits is difficult but not impossible for antifuse. However, once the location of anti-fuse is extracted the stored bit can be probed. 
Moreover, all the OTP requires higher breakdown voltage and a large peripheral circuit which introduce area overhead and power consumption.

Other conventional examples of NVMs are EEPROM and Flash memories.
Each EEPROM cell has two transistors - a floating gate or storage transistor and a select transistor. 
The storage transistor has a floating gate which traps the electrons. 
A Flash cell only has the floating gate transistor and use the same logic storage mechanism as EEPROM. 
Since both memory technologies use stored charges in the floating gate for storing the bit values, any attempt to image the memory cell with SEM or TEM can disturb the charges distribution and possibly erase the memory content.
Therefore, reverse engineering of such NVMs has always been considered as a challenging task; even after the recent advancements in FA tools.
Nardi et al.~\cite{de2005oxide} solved the challenge of maintaining the value of stored charge by accessing the memory from the back-side of IC.
Once an attacker gets access to the floating gate of EEPROM/Flash, she can use scanning Kelvin probe microscopy (SKPM), scanning probe microscopy (SPM), passive voltage contrast (PVC) or scanning capacitance microscopy (SCM) for extracting the stored value in the EEPROM/Flash~\cite{de2005oxide, courbon2016reverse}. 
However, the security of the 3D Flash chips (see 3D NAND flash cells in Fig.~\ref{fig:3D_NAND}) have yet to be investigated. 
In the 3D flash technology, the memory cells, previously organized horizontally, are now stacked vertically and connected with pillar and channels.
Although such orientation requires further precaution during polishing the back-side of the chip and PVC analysis, the reverse engineering of 3D NAND memory is, in principle, still possible. 

Physical unclonable functions (PUFs), as other possible candidates for secure key-storage, was developed to generate keys from intrinsic properties of the device~\cite{tajik2015laser}. 
Although PUF has been assumed to be tamper-evident against physical attacks, they have demonstrated vulnerabilities against several non- and semi-invasive attacks, like photonic emission analysis and laser fault injection~\cite{ tajik2015laser} 
Furthermore, the response of PUF differs for each chip due to process variation which makes it incompatible for ASIC design, where the same mask would be used for fabricating all the chip in the same batch.
On the other hand, storing the key value in the battery-backed RAM also does not add any significant security feature to the key-storage as they can be read out through optical attacks, such as thermal laser stimulation (TLS)~\cite{lohrke2018key}.

Data remanence in key-storage like NVM and RAM is another class of vulnerability for all key-storage elements.
Data remanence is the residual physical representation (e.g., the trapped charge or voltage) of the data that has been erased from the memory during a tampering attack or regular operation of the chip.
A tamper-sensor enclosure can initiate the erasure procedure for memory if the tampering event is detected.
The sensor connects the memory to the ground to zeroized the stored data.
However, due to data remanence effect, an attacker can exploit the residual property of the memory to extract the content of the memory. 
The data remanence vulnerability occurs when data retention time exceeds the time required by a malicious entity to read out or dump the stored value in another memory location.
Consequently, the protection mechanism can be defeated~\cite{skorobogatov2018hardware}.

\subsection{Vulnerabilities of the Interconnects}\label{sec:interconnect_vul}

Sensitive information transmitted on wires in ICs can be physically extracted using contact-based electrical probing attack~\cite{wang2016probing}.
In this type of attack, the chip's wires are contacted by a probe, and as a result, the signal carried by the wires can be read out when the chip is functioning.
Therefore, electrical probing is considered as a \textit{contact-based} method for extracting the assets in the chip. 
Electrical probing attacks can be classified into frontside probing, which is carried out through the passivation layer and upper metal layers, and  back-side probing, which is mounted through the silicon substrate.

\begin{figure}[!t]
\centering
\includegraphics[width=0.35\textwidth]{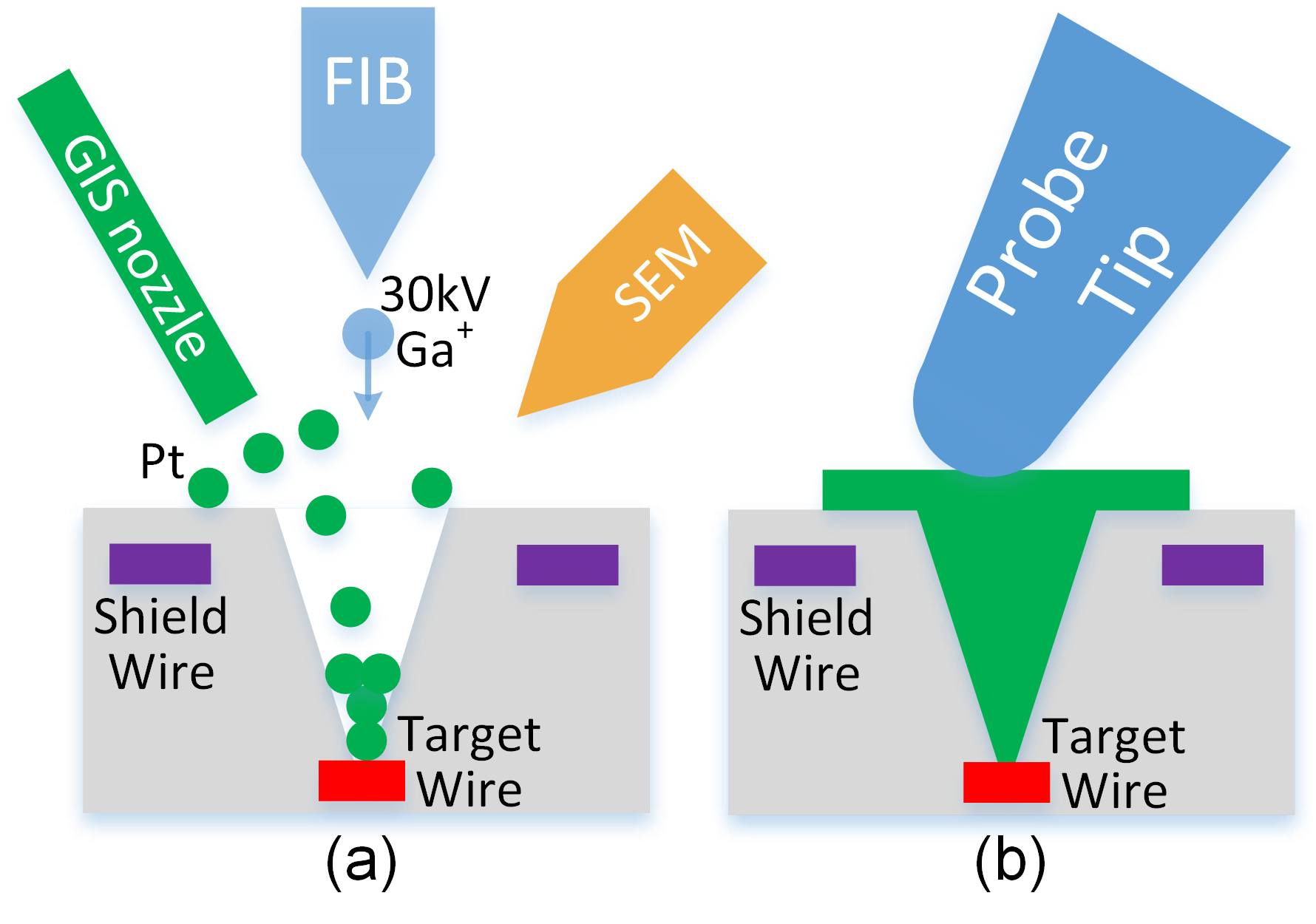}
\caption {(a) FIB deposits Platinum in the milling cavity to build a conducting path (green) from the target wire; (b) The deposited conducting path serves as a electrical pad for the probe contact~\cite{Wang2019}.}
\label{fig:fib_milling}
\end{figure}

Due to the large size of probes in comparison to the size of metals' width and available space between wires, the frontside electrical probing is a challenging task.
To overcome these limitations, attackers usually deploy focused ion beam (FIB), which is a powerful tool commonly used in the testing, development, and editing of ICs with nanoscale precision, to mill a narrow cavity, get access to the target wire on lower metal layers, and build a conducting path without damaging upper metal layers as shown in Fig.~\ref{fig:fib_milling}.
Modern FIB systems, such as ZEISS ORION NanoFab, can edit out obstructing circuitry with a 5~nm precision. 
FIB aspect ratio is a key feature of FIB's capability, which is defined as the ratio between the depth and diameter of the milling cavity.
Thus, the higher of the FIB aspect ratio, the thinner of the milling cavity, the less probability to damage signal wires on the chip, and the higher success rate to extract wire values.

Some high-security level chips, such as smart cards, may have shield-like mechanisms to protect the chip against frontside probing attacks.
However, this type of countermeasure may still be compromised by bypass and reroute attacks~\cite{wang2016probing} using advanced FIBs.
In the case of bypass attacks, the attacker can utilize the limited space between shield wires to approach lower target wires without hurting the adjacent shield wires using high aspect ratio FIB.
For reroute attacks, on the other hand, the attacker can build a copy path between two equipotential points on shield wires using FIB's deposition capability, so the original path between these two equipotential points can be cut at will.
As a result, even shielding cannot provide adequate security protection and it can still be vulnerable to sophisticated attackers equipped with advanced FIB systems.
The electrical probing attack can be mounted from the backside of the IC as well\cite{helfmeier2013breaking}.
In this case, the silicon substrate on the backside of the chip is penetrated to create access to the lower metal layers.
Therefore, while reaching sensitive wires on the lower metal layers is challenging through frontside attacks, they can be accessed through the backside where there are little to no protection mechanisms.

\begin{figure*}[!tbh]
        \centering
        \begin{subfigure}[b]{0.25\textwidth}
        \centering
                \includegraphics [width=\textwidth]{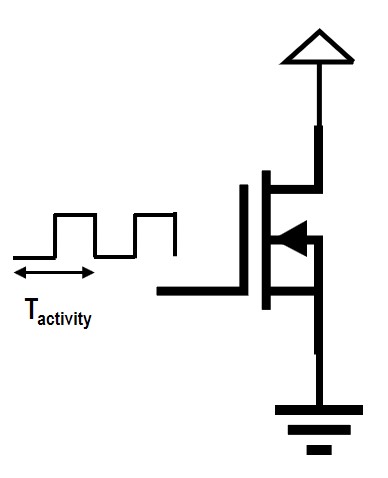}
                \caption{}
                \label{fig:EOFM_CIRCUIT}
        \end{subfigure}%
        \hspace{3pt}
        \begin{subfigure}[b]{0.4\textwidth}
        \centering
                \includegraphics [width=\textwidth]{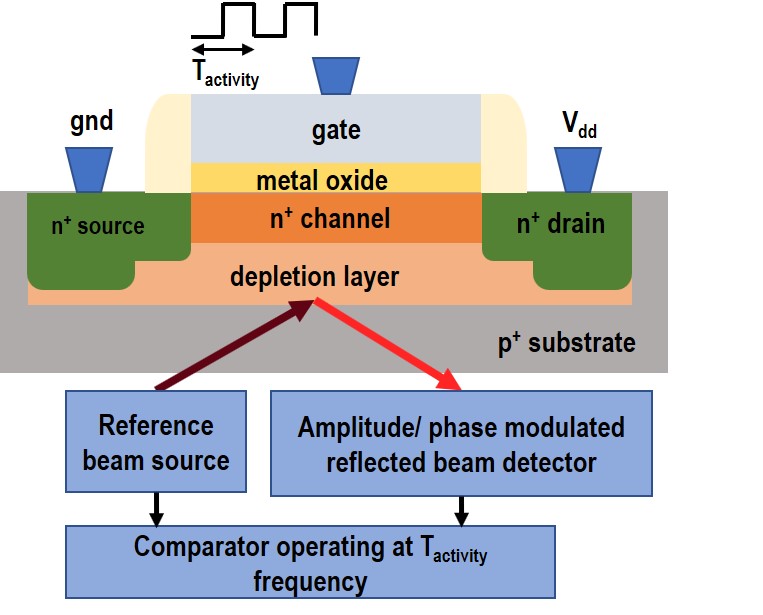}
                \caption{}
                \label{fig:EOFM_PRINCIPLE}
        \end{subfigure}
        \hspace{1pt}
        \begin{subfigure}[b]{0.3\textwidth}
        \centering
                \includegraphics [width=\textwidth]{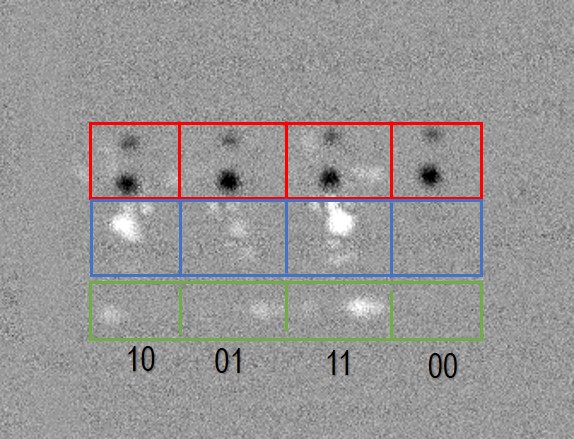}
                \caption{}
                \label{fig:EOFM_REG}
        \end{subfigure}
\caption{(a) The input signal connected to the gate terminal of an n-MOSfet operating at T\textsubscript{activity} frequency; (b) Reference beam got modulated due to the activity of the transistor. The modulated reflected beam is compared and filtered at the same frequency at the gate is operating; (c) EOFM activity measurement of a 8-bit register. The black dots in red rectangles represent the clock activity, white dots in blue rectangles represent flip-flop activity and white dots in green rectangles represent the output buffer activity. The stored value in each register is mentioned at the bottom of the output buffer.  }
\label{fig:EOFM}
\end{figure*}

\subsection{Vulnerabilities of the Key-delivery Unit}\label{sec:key_delivery_vul}
Similar to contact-based methods, the contactless optical probing~\cite{tajik2017power} techniques can impose the threat of exposing security-sensitive information to an adversary, e.g., the key value in logic locking schemes.
Optical probing is a semi/non-invasive chip debugging method, which enables the probing of the volatile and on-die-only values of key-registers and key-gates at run-time.
In modern ICs, multiple interconnect layers at the frontside of the chip obstruct the optical path from the transistor. 
On the contrary, no such protection is available on the backside of the device. 
Hence, attacking the logic locking and FSM using optical probing is more convenient if conducted from the backside.

In optical probing the chip must be operation. 
Therefore, the selection of sample preparation method for the DUA depends on the packaging, i.e., non-flip or flip chip packaging technique.  
In non-flip chips, the die backside can be accessed by decapsulating the packaging. 
Such challenges can be avoided if the DUA is in a flip-chip package.
The silicon substrate in a flip-chip package is usually covered with a heat-sink which can be removed easily using a lab knife and hotplate~\cite{cryptoeprint:2019:719}. 
Once the chip is decapsulated, the device receives a global polishing to increase the resolution for back-side FIBing and electrical probing. 
In flip-chip, such polishing is not necessary for optical probing, and therefore, optical probing can be considered as a non-invasive physical attack which makes such attack more attractive to an adversary~\cite{cryptoeprint:2019:719}.
Besides, in the case of optical probing, the spatial resolution can be increased if the adversary has access to solid immersion lens (SIL).     

To attack the key-delivery unit using optical probing, an adversary requires access to a laser scanning microscope, which is available in advanced FA labs.  
Since silicon is transparent to near-infrared (NIR) light source, the activity in the die can be measured using electro-optical frequency (EOFM) and electro-optical probing (EOP)~\cite{tajik2017power}. These two methods are major optical techniques used for debugging nanoscale transistors.
In both EOP and EOFM, the incident photons with NIR wavelength pass through the back-side of silicon substrate which leads to partial absorption and reflection at interfaces like back-side silicon and active region or first metal layer interconnect.
In the case of EOP, the electrical signal at a node modulates the amplitude and phase of reflected light. 
The modulated light is fed to an optical detector and compared with the reference NIR wavelength laser beam (see Fig.~\ref{fig:EOFM_PRINCIPLE}). 
As the modulation of the reflected beam signal is small, a sufficient signal-to-noise ratio is acquired through running the signal in a certain trigger frequency (T\textsubscript{activity} in Fig.~\ref{fig:EOFM_CIRCUIT} and Fig.~\ref{fig:EOFM_PRINCIPLE}) and measuring the signal. 
In EOFM, a laser scans the region of interest (ROI) on the device under attack and feeds the detected signal from laser reflected signal into a spectrum analyzer acting as a narrow band frequency filter, for example in the Fig.~\ref{fig:EOFM} the frequency of narrow bandpass filter of the spectrum analyzer is T\textsubscript{activity}. 
The output from spectrum analyzer is mapped in a 2D image using grayscale or false color representation \cite{tajik2017power}. 
Analyzing the output from EOP or EOFM, the data stored in a node is extracted.  
The EOFM activity of an 8-bit register measured at two different frequency -- clock frequency and T\textsubscript{activity}, and stored value in the 8-bit register is shown in Fig.~\ref{fig:EOFM_REG}. 
Hence, an adversary can probe the data stored in the registers from the backside of the chip die without using the invasive methods like FIB.  
   
A malicious entity can always use advanced reverse engineering tools to extract the gate-level netlist of the chip. 
Access to gate-level netlist enables the intruder to dig deeper in the chip design and localize the key-gates and key-delivery unit or the interconnects carrying the locking key of the chip. 
Therefore, by learning the operating frequency for the key-delivery unit and using EOFM, an attacker can probe different key-carrying elements like key-gates, key-registers or key-management logic and learn the locking key~\cite{cryptoeprint:2019:719}. 
Hence, optical probing is a direr threat for logic obfuscation as this method can extract the locking key in a contactless manner; without using invasive methods, like FIBing or circuit edit, and contact-based method, like electrical probing. 



\subsection{Vulnerabilities of the DFT} \label{SEC:dft_vulnerability}
Jeopardized by the worldwide IC supply chain, scan infrastructure can be used to assist non-invasive attacks, thereby compromising security. The exposed scan chains may leak critical information such as intellectual property (IP) or secret keys to the attackers, 
which can be carried out by any entity within the IC supply chain. Hence practical solutions are needed to protect ICs against scan-based side-channel attacks \cite{dworak2015call}. 
In the last decade, there have been a number of scan-based attacks on various cryptosystems. In \cite{mukhopadhyay2005cryptoscan}, the risk of scan-based attack is presented as a general threat to a stream cipher. To obtain critical information, the attackers can ascertain the internal structure of the scan chain by running encryption in normal mode and then switching to test mode. \cite{yang2006secure} 
have successfully uncovered scan-based attacks on the dedicated hardware implementation of the Data Encryption Standard (DES), Elliptic Curve Crypto-systems (ECC), Advanced Encryption Standard (AES), and RSA. Since scan chains directly reveal the internal state of the logic blocks, attackers can use them to perform IP piracy  
With the knowledge of the design,  
attackers can also control the chip without authorization by scanning illegal values into the system status registers to disrupt the chip. In light of these threats, ensuring scan security has become a great concern to the industry, and various countermeasures have been proposed which are summarized in Table \ref{scan_table}. A detail discussion of these threats and existing countermeasures are discussed below.

\begin{itemize}[leftmargin=*]
    \item Differential Attack and Defense: The differential attack \cite{rolt2013novel} is based on applying challenge pairs, running the crypto algorithm, and comparing the outputs to extract the key. This attack has been facilitated using scan chain due to added controllability and observability. Through switching from functional mode to test mode, the attacker can identify key flip-flops from the scan chain. 
    Then, the key can be recovered through the already constructed correlation among input pairs, key flip-flops, and key \cite{rolt2013novel}. 
    The most direct solution to refrain from differential attack is to defuse the poly-silicon fuses connecting the scan-in or scan-enable ports \cite{kommerling1999design}; however, this prohibits in-field testing which is a must in advanced ICs.
    Some test mode protection techniques have been proposed\cite{hely2006secure,chiu2008ieee} which attempt to reset the data registers when the chip is switched to test mode and wrap the non-volatile memories. However, test mode only differential attacks\cite{saeed2014test} successfully extracted the key.
    
    \item Advanced Industrial DFT Techniques: On-chip compression, X-tolerance, and X-masking are considered natural barriers to scan-based attacks \cite{liu2007effects}. However, the compression bypassing mode is always kept for the sake of debugging and diagnosis. Recently some attacks have been made even in the presence of on-chip compression \cite{saeed2014test}, X-masking \cite{da2012advanced}, and X-tolerance \cite{darolt2011scan}.
    
    \item Scan Interface Encryption: In addition to the on-chip compression used in advanced DFT structures, scan chain encryption has been developed as countermeasures. In \cite{da2017scan}, the scan patterns/responses are decrypted/encrypted at each scan input/output, respectively, which is conducted by highly efficient and secure block cipher at each scan port.
    But these countermeasures are defeated by resetting attack\cite{sengar2007secured} and flushing attack\cite{atobe2012dynamically}. By resetting the scan cells or flushing the scan chain with the known patterns, the fixed inverted bits \cite{sengar2007secured} and modified bits \cite{atobe2012dynamically} in the obfuscated scan chain can be identified so that the plaintext can be deciphered.
    
    \item Partial Scan: The secure scan architectures presented in \cite{yang2006secure} exclude flip-flops containing sensitive information from the scan chain. However, only part of the scan chain cells can be protected. Besides, defects in the excluded registers cannot be detected, which decreases the test coverage and potentially impacting yield.
    
    \item Obfuscated Scan: In \cite{sengar2007secured, atobe2012dynamically, lee2006low, razzaq2011sstkr, paul2007vim}, dummy flip-flops or other obfuscation logic (i.e., inverters, XOR gates, etc.) have been inserted into the scan chain to randomize scan outputs. A scan chain access authorization process usually controls obfuscation. The scan out responses are determined by the test authentication status. However, some obfuscation logic inserted into the scan chain are not robust against reset or flushing attacks \cite{sengar2007secured, atobe2012dynamically}. More importantly, the scan authorization key bits hidden in the test patterns are usually easy to locate \cite{lee2006low, razzaq2011sstkr, paul2007vim}. Furthermore, the authentication key bit flipping would make scan out vectors differ, while a non-key bit would not. This would significantly reduce the difficulty of identifying the key bits and becomes vulnerable to bit-role identification attack\cite{wang2017secure}.
    
    \item Scan Chain Reordering: In \cite{hely2004scan}, the order of scan cells is dynamically reconfigured by an unpredictable scrambler, which increases the routing overhead significantly. In \cite{lee2007securing}, each scan chain is divided into several segments, and then the test controller determines the segments’ scanning out sequence. In \cite{mukhopadhyay2005cryptoscan, sengar2007secured}, scan tree architecture is applied to reorder the scan chains. However, these methods still could not defend against a differential attack \cite{saeed2014test}, and require significant change to the DFT flow.
    
    \item Combinational Function Recovery Attack: Since the scan chains unfold the sequential logic as combinational and directly reveal the internal states of the circuit, extracting design information from them has become easier. Thus, the device’s functionality can be reverse engineered \cite{azriel2016exploiting}.
    
      \item Oracle-guided Attacks: While logic locking can be an effective technique to establish trust among different entities of the IC supply chain, it has not seen application due to its lack of attack resiliency. 
      The logic locking is proved to be vulnerable against Oracle-guided attacks. 
      In Oracle-guided attacks, the attacker has access to an unlocked or functional chip.  
A functional chip carries the key value in the key-storage element. 
Therefore, such an IC can generate the correct output for any input pattern and the attacker can make use of the correct input/output pairs to rule out incorrect keys and extract the correct obfuscation key.
      For example, most logic obfuscation techniques are vulnerable to Boolean satisfiability (SAT)-based oracle-guided attack, key-sensitizing attack \cite{rajendran2012security} and EPIC attack \cite{plaza2014protecting}. 
      The key sensitizing attack utilizes automatic test pattern generation (ATPG) tool to propagate the effect of a key gate to a primary output. 
      SAT attack \cite{subramanyan2015evaluating} breaks most combinational logic obfuscation techniques in a short matter of time by finding distinguishing input patterns (DIPs).
      DIPs rule out incorrect keys utilizing the output corruptibility of the miter circuit constructed using locked design and activated design. 
      For sequential designs, it is assumed that an IC's internal states can be accessed and controlled via scan chains to read/write the value of the flip-flops. To resist SAT attack, several SAT-resistant logic obfuscation techniques have been proposed- SARLock\cite{yasin2016sarlock}, Anti-SAT\cite{xie2018anti} and SFLL \cite{yasin2017provably}. SARLock and Anti-SAT resists SAT attack by increasing the number of required distinguishing input patterns (DIPs), thus exploiting a point function to corrupt the output of the design for all the incorrect keys. While these two SAT resistant techniques are strong enough to withstand the power of oracle-guided attacks, they are vulnerable to Bypass attack\cite{xu2017novel}, SPS attack\cite{yasin2017removal}, and AppSAT \cite{shamsi2017appsat} attack. SFLL\cite{yasin2017provably} technique strips some of the functionality of the original design and hides it in the form of a secret key. Once correct secret key is applied, original functionality of the design is restored. SFLL was briefly considered the state-of-the-art SAT resistant logic obfuscation technique. Then a recent functional analysis attack (FALL) \cite{sirone2019functional}  was proposed that uses structural and functional analyses on the locked design to identify the locking key, without even having access to an oracle. EPIC attack \cite{plaza2014protecting} uses a hill-climbing search based algorithm that monitors test response to guess the secret key. The attack tries to reach zero hamming distance between the test response of the activated IC and the encrypted circuit by flipping the individual bits of the initial key guess if the flip reduces the hamming distance.
      
\end{itemize}

From the above discussion it is apparent that none of the existing countermeasures can provide full protection against attacks that exploit scan infrastructure. For example, most countermeasures targeting scan-based side-channel attacks, do not consider protecting against IP Piracy, over-production, tampering and counterfeiting. A dynamic scan chain obfuscation technique\cite{wang2017secure} has been suggested for protecting IPs against most of the scan-based attacks discussed above by dynamically changing scan obfuscation key and scrambling scan-in patterns and scan-out responses. But this countermeasure does not consider the threat model of oracle-guided attacks e.g., SAT attack \cite{subramanyan2015evaluating}. Vulnerabilities of these attacks are further discussed in Sect.~\ref{sec:obfuscation_technique_vul}. Hence, developing such countermeasure is    necessary that can protect its secret against not only scan-based side-channel attacks abut also scan facilitated oracle-guided attacks.

\begin{table}[t]
\centering
\caption{Scan-based Attack and Countermeasures}
\label{scan_table}
\begin{tabular}{|l|l|l|}
\hline
\multicolumn{1}{|c|}{\bf Attacks} & \multicolumn{1}{c|}{\bf Exploits} & \multicolumn{1}{c|}{\begin{tabular}[c]{@{}c@{}}\bf Existing\\ Countermeasures\end{tabular}} \\ \hline
\begin{tabular}[c]{@{}l@{}}Differential\cite{rolt2013novel}/Test\\ mode only Attack\cite{ali2013new}\end{tabular} & Internal States & \begin{tabular}[c]{@{}l@{}}Scan encryption\cite{da2017scan},\\ DOS\cite{wang2017secure}\end{tabular} \\ \hline
Resetting Attack\cite{sengar2007secured} & \multirow{3}{*}{Internal Secrets} & \multirow{3}{*}{\begin{tabular}[c]{@{}l@{}}LCSS\cite{lee2006low}, DOS\cite{wang2017secure},\\ Lock \& Key\cite{lee2005securing},\\ Scan encryption\cite{da2017scan}\end{tabular}} \\ \cline{1-1}
Flushing Attack\cite{atobe2012dynamically} &  &  \\ \cline{1-1}
Bit-role Identification &  &  \\ \hline
\begin{tabular}[c]{@{}l@{}}Combinational Function\\ Recovery\cite{azriel2016exploiting}\end{tabular} & Functionality &  DOS\cite{wang2017secure}\\ \hline
SAT Attack\cite{subramanyan2015evaluating} & Obfuscation Key & \begin{tabular}[c]{@{}l@{}}SARLock\cite{yasin2016sarlock},\\ Anti-SAT\cite{xie2018anti},\\ SFLL\cite{yasin2017provably}\end{tabular} \\ \hline
\end{tabular}
\end{table}

\begin{figure*}[t]
\centering
\includegraphics[width=0.95\textwidth]{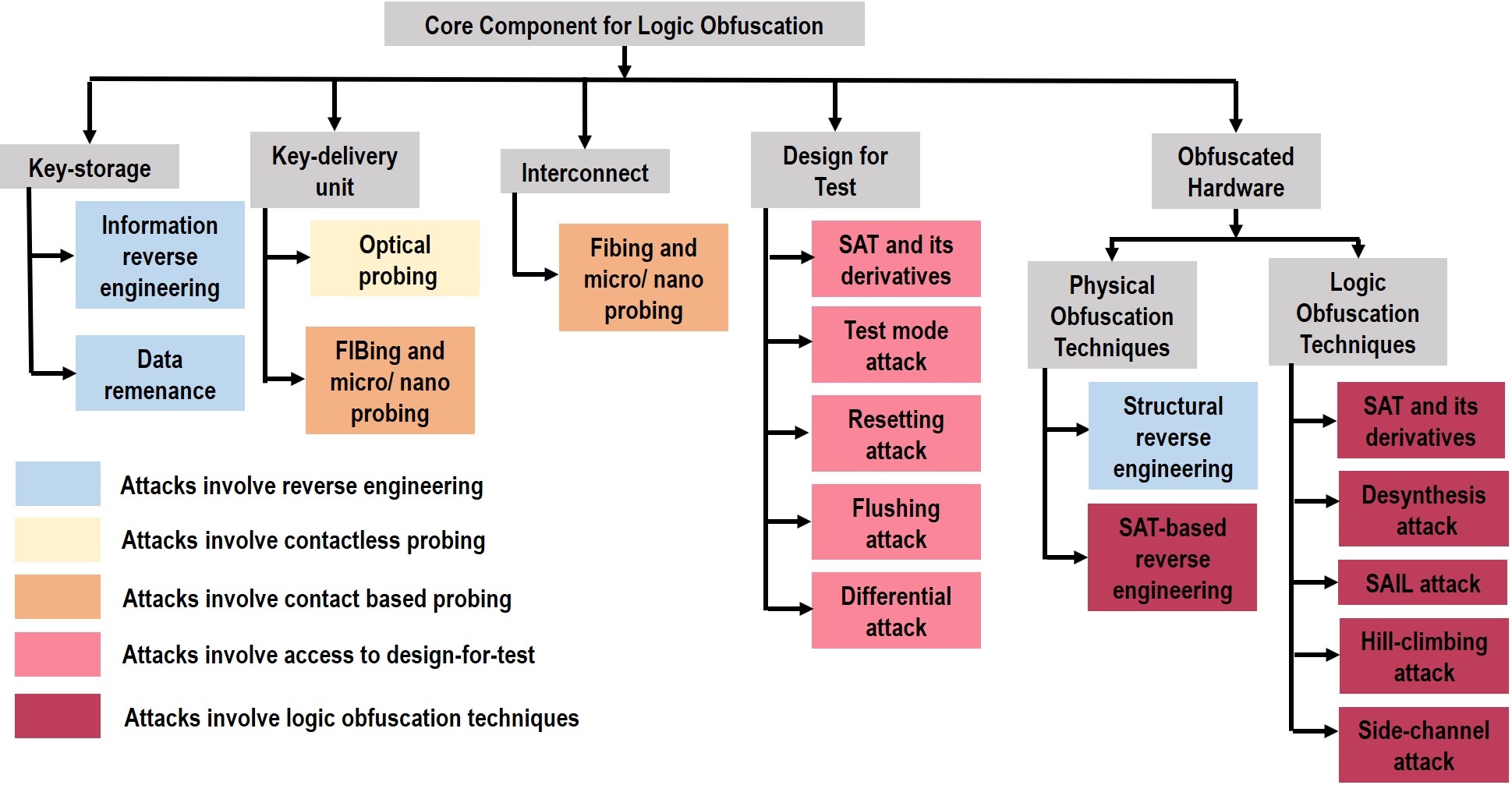}
\caption{Attack methods for the core element in a logic locked chip.}
\label{fig:framework}
\end{figure*}

  \subsection{Vulnerabilities of the Obfuscated Hardware}\label{sec:obfuscation_technique_vul}
  The source of the vulnerabilities for obfuscated hardware lies in the techniques used for obscuring the functionality and layout of the chip. 
  Any shortcoming in the security of obfuscation techniques weakens the security of obfuscating key  as well as all the assets in the chip.
  Therefore, we analyze the vulnerability of logic locking and physical obfuscation techniques in detail.  
  
  \subsubsection{Vulnerabilities of Logic Locking Techniques}
  In the past decade, there has been a number of attacks proposed to retrieve the key from the logic locked circuit. 
  The attacks available in the literature can be classified into two classes -- Oracle-guided attacks and Oracle-less attacks. 
  Different Oracle-guided attacks are described in detail in Sect.~\ref{SEC:dft_vulnerability}. 
Along the aforementioned Oracle-guided attacks, side-channel information like differential power analysis and test data can be used to learn the key value in a locked chip. 
Over the past several years, the security community has focused on assessing the vulnerabilities due to Oracle-guided attacks. 
While protecting the structural obfuscation from the above-mentioned attacks received so much attention, unfortunately, no evaluation has been performed to find the information that can be extracted from the netlist alone. 
The change due to logic locking in the netlist is local, i.e., the key-gates combine with the logic elements in the netlist to transform a new structure. 
Such structure can also be identified if the adversary has prior knowledge about the synthesis tools. 
Therefore, in desysnthesis attack \cite{massad2017logic}, authors have proposed, re-synthesizing the locked netlist with a random key and then using hill climbing search to find the key value yields the maximum similarity between the locked netlist and re-synthesized netlist. 
Using machine learning techniques, it is also possible to revert the locked circuit into the pre-synthesis version of the design and retrieve the original design and functionality of the chip~\cite{chakraborty2018sail}. 

\subsubsection{Physical Vulnerabilities to Reverse Engineering the Obfuscated Hardware}
Physical obfuscation mainly focuses on preventing the reverse engineer from stripping the ICs layer by layer and extracting gate-level for duplicating a netlist without authorization of the IP holder. 
Shrinking the device dimension was never an issue for reverse engineering. 
Continuous improvement and automation in FA tools along with the netlist extraction software, such as Pix2Net, Degate, etc. 
always proved to be successful against smaller node technologies like 14mm . 
The reverse engineering software use image processing techniques to identify the functionality of the gates. 
In order to thwart automated image processing based reverse engineering, several subtle obfuscation techniques like gate camouflaging, dummy contacts, dummy interconnects , filler cells, variation in doping concentration have been proposed \cite{rajendran2013security, becker2014stealthy}. 
However, layout obfuscation methods can be detected using advanced imaging tools like PVC, SEM or dynamic optical beam induced current circuit analysis (DOCA) \cite{pix2net}. 
Using PVC or varying the beam voltage of an SEM, a reverse engineer can distinguish between the active cell and filler cells due to variation in doping concentration \cite{vijayakumar2017physical}. 

The aforementioned camouflaging techniques are not only vulnerable to failure analysis tools, but they are also vulnerable to several attack methods, as for example SAT attack, brute force attack, and behavior analysis. 
An adversary can isolate the camouflage gates and sensitize the output of the gate using input pattern to resolve the functionality of the gate using the brute force attack~\cite{rajendran2013security}. 
Again, the adversary can perform behavior matching against a library of components with known functionalists to expose the functionality. 
SAT-based de-camouflaging and removal attacks can also debunk the gate level camouflaging  \cite{yu2017incremental}.  
\subsection{Security Breach Through Hardware Trojan Insertion}
Device assets such as the locking key should be protected by hardware. 
The hardware contains physical countermeasures against several physical attacks, tampering, side-channel analysis and probing in particular.   
The aforementioned protection imposes a significant barrier to attackers thus implicitly providing a basic level of protection against key extraction. However, an untrusted foundry can intentionally introduce side-channel leakage by inserting hardware Trojan in the design, in a similar fashion described in \cite{lin2009trojan} for the key of cryptomodule. 
Identifying the location of the key-storage elements and the key-delivery unit and implementing a Trojan to facilitate the side-channel analysis can empirically serve the purpose. 
Hence, the possibility of a security breach due to the presence of hardware Trojan into the design cannot be ignored.

\subsection{Summary of the Vulnerabilities of the Core Elements}
Each of the core components described in Sec. \ref{Sec:corecomponent} acts as a link in the web of logic locking to defend the chip design form IP piracy and violation of root-of-trust. 
On the basis of the above discussion, the attack methods for breaking into the core components of an obfuscated chip and tamper its security can be categorized in five classes; 
\begin{enumerate}
\item attacks that involve either structural or information reverse engineering methods, 
\item attacks that involve contactless probing methods like optical probing. In such methods, no direct contact with the transistors is required for extracting the secret data like locking key, 
\item attacks that involve contact-based probing methods like electrical probing, 
\item  attacks that involve access to design-for-test structure such as scan chain, and 
\item attacks on logic obfuscation techniques, for example, SAT and SAIL attack. 
\end{enumerate}
 Fig.~\ref{fig:framework} summarize the vulnerabilities of the core components based on the above-mentioned five attack categories.



\section{Threat Model Analysis: Security Threats in IC Supply Chain } \label{sec:supply_chain}
In this section, the security and trust issues in the supply chain, the stake holders, and the threat analysis for potential adversaries are discussed.

\begin{figure*}[t]
\centering
\includegraphics[width=0.85\textwidth]{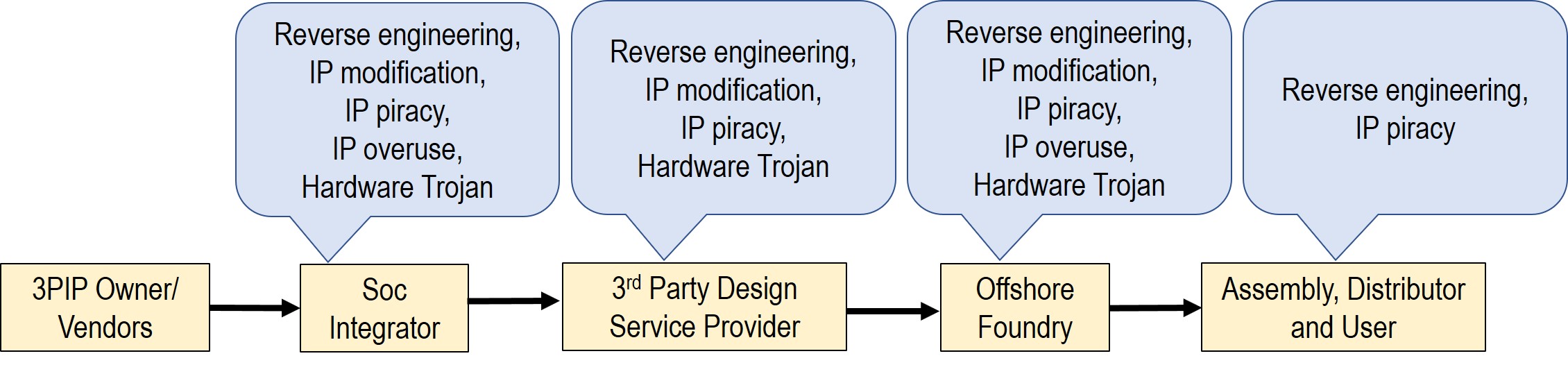}
\caption{Stake holders and corresponding IP threats in the horizontal supply chain. }
\label{fig:supply_threat}
\end{figure*}

\subsection{Vulnerability Analysis in Supply Chain of SoC}
\label{sec:supply_chain_1}
In the last decade, the SoC supply chain has shifted to a horizontal business model. 
In the horizontal model, several stakeholders are involved in the manufacturing steps and supply chain of the SoC (Fig.~\ref{fig:supply_threat}). 
Usually, OCM starts the design process by acquiring the IP which is developed in-house or purchased from \textit{third-party IP vendors} (3PIP Vendors). 
Later, the SoC designer incorporates the in-house developed and procured 3PIPs to generate the RTL specification of the whole SoC. 
The SoC integrator synthesizes the RTL description into a gate-level netlist using computer-aided design (CAD) tool, for example, Design Compiler from Synopsys. 
The gate-level netlist then goes through formal equivalence checking to verify that the netlist is functionally equivalent to the RTL representation. 
Moreover, the gate-level netlist is also verified to check if the design meets timing, power, and area requirements. 
Thereafter, the SoC integrator integrates the DFT structure to enable the IC to be thoroughly tested during fabrication, package assembly, and in the field operation to ensure its correct functionality.
Due to aggressive time-to-market, design houses may outsource some portion of the design, e.g., DFT insertion, physical layout design, to \textit{third-party design service providers} and receive final GDS from them.  
In the past two decades, most design houses have become fabless. 
Therefore, they fabricate their products in \textit{third-party offshore foundries}. 
In this process, the SoC designe house can enjoy the state-of-the-art fabrication technologies, however, at the cost of reduced trust in the manufacturing process (product integrity will be in doubt). 
After fabrication, the offshore foundry sends tested wafers to the assembly line to cut the wafers into die, and package the good ones to produce chips. 
After these processes are done, assembly performs structural tests to find defects in the chip that could be introduced during the assembly process. 
After performing these tests, the chips without defects are shipped to the distributors, or the system integrator. 
The distributors sell these ICs in the market.
With all these discussions we can summarize that IC design flow encompasses entities that design their own chips (fabless design houses), entities that offer design services to other firms (third-party design service providers or IP vendors), entities that offer fabrication facilities (offshore foundries), and entities that design and manufactures their chips in-house~\cite{fuller2014chip}.
Different stakeholders in the supply chain have different motivations for IP infringements, therefore, introduce different vulnerabilities in the supply chain, as shown in Fig.~\ref{fig:supply_threat}

\subsection{Potential Adversaries} \label{sec:threat_model}

The objective, assets and capabilities available to an attacker influence the vulnerabilities that she might be interested to exploit. 
As shown in Fig.~\ref{fig:supply_threat}, the untrusted foundry, SoC integrator, third-party design service provider, and end users can be identified as the potential antagonist against logic obfuscation.

\subsubsection{Foundry}
The combinational logic locking and FSM locking consider an offshore foundry as the primary source of threat in the supply chain \cite{harpoon, subramanyan2015evaluating, xu2017novel}.
Since the foundry has access to the GDS II file which they use to develop the costly mask for chip fabrication, an untrusted foundry is a major suspect for IP infringement. 
In addition, the attacker also can obtain an activated chip from the open market, a malicious insider in a trusted entity in the supply chain, or from a fielded system. The capability of each foundry also includes access to the state-of-the-art FA tools and reverse engineering capabilities. 
Access to DFT structure for detecting and analyzing the failure in the die is another asset available to the foundry.  
Access to aforementioned capabilities enables the foundry to reverse engineer the chip and localize the key-storage element, interconnect, key-delivery unit, key-gates, and DFT distribution to bypass the security of the obfuscated design. 
Consequently, the implementation of the circuit is crystal-clear to the foundry. 

\begin{figure}[t]
        \centering
        \begin{subfigure}[b]{0.48\textwidth}
                \includegraphics [width=\textwidth]{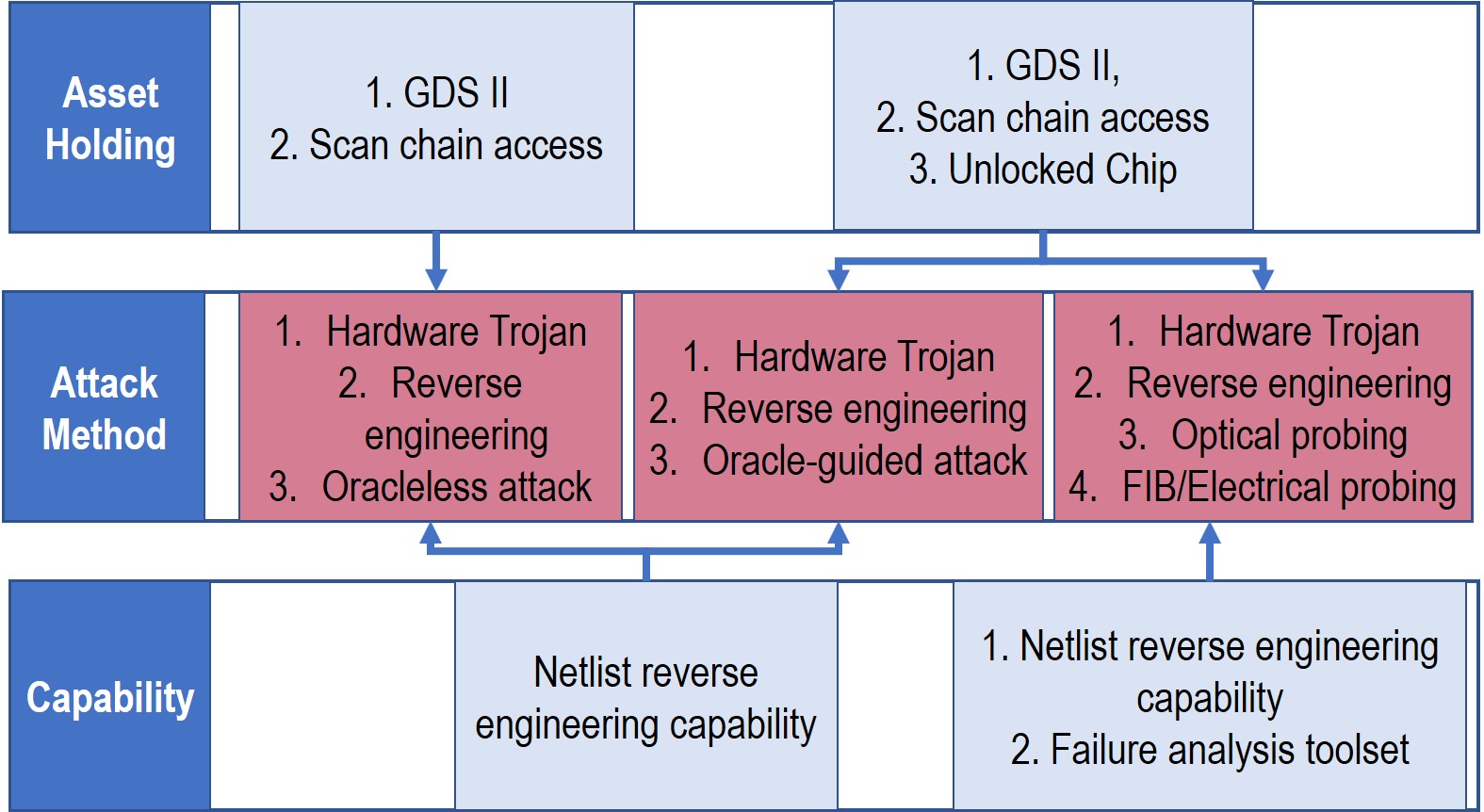}
                \caption{}
                \label{fig:foundry_model}
        \end{subfigure}
        \hspace{3pt}
        \begin{subfigure}[b]{0.47\textwidth}
                \includegraphics [width=\textwidth]{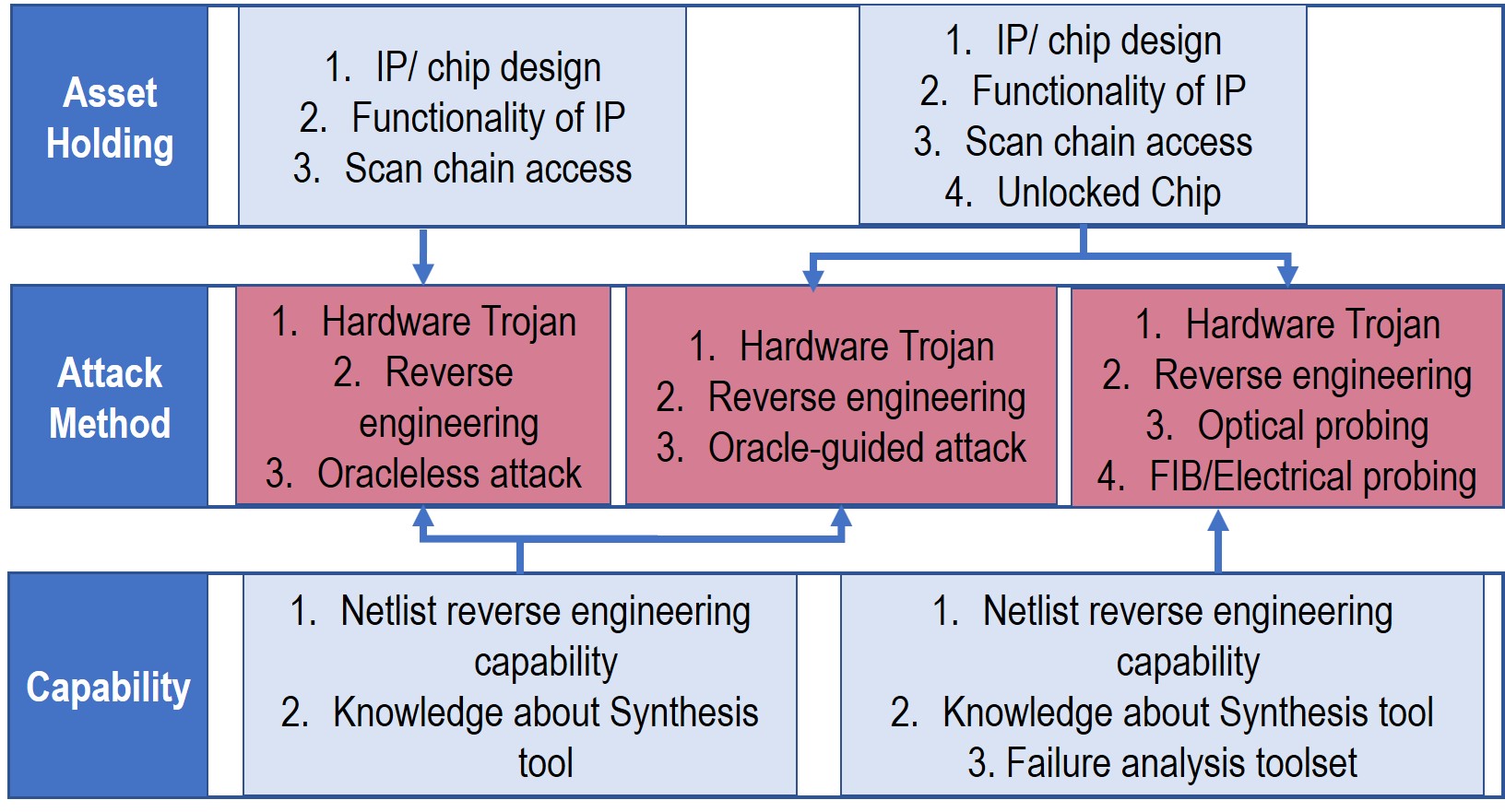}
                \caption{}
                \label{fig:socdesigner_model}
        \end{subfigure}
        \hspace{1pt}
         \begin{subfigure}[b]{0.26\textwidth}
                \includegraphics [width=\textwidth]{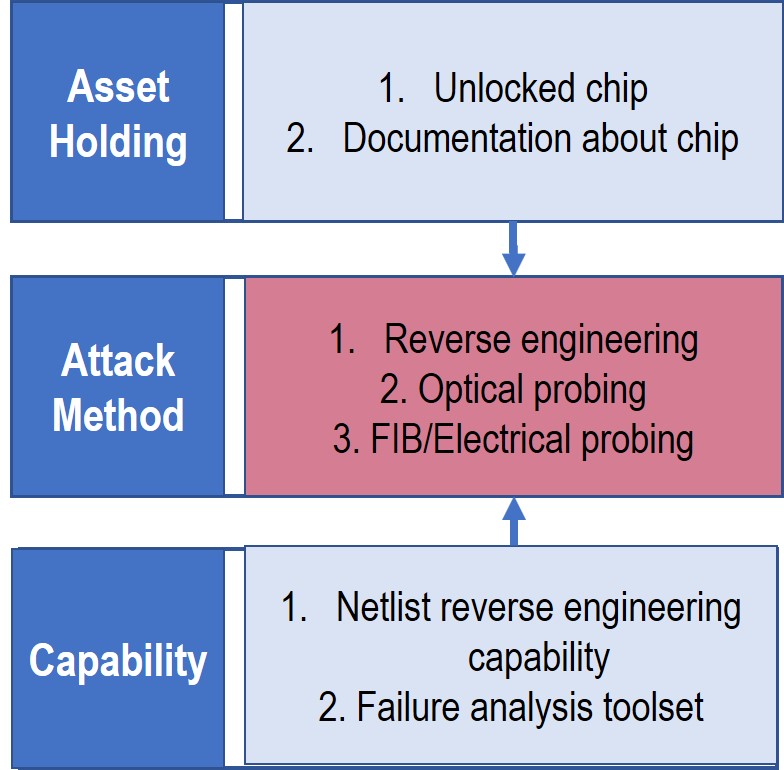}
                \caption{}
                \label{fig:enduser_model}
        \end{subfigure}
\caption{The threat model depending on asset and capability available to different untrusted entity in the supply chain -- a) threat model for the untrusted foundry, b) threat model for the untrusted 3rd party design service provider and the SoC designer, c) threat model for the end user.}
\label{fig:threat_model}
\end{figure}

The objective of a rogue foundry is overbuilding and selling the chip in the open market. 
The adversary can also locate any specific IP from the design and learn about the implementation and functionality of that IP for hardware Trojan insertion or IP piracy. 
Depending on the objective of attack and obfuscation technique implemented in the design, a malevolent foundry can select its attack methodology. 
As the foundry can learn about the location of key-gates and key-delivery unit; applying FA methods like optical and electrical probing for extracting the key value of the key-gate is more convenient for the attacker \cite{cryptoeprint:2019:719, rahman2018physical}.
However, foundry can perform black box analysis of structural obfuscated chip and exploit the Oracle-guided (for example, SAT, bypass, and SPS attacks) and Oracle-less (for instance SAIL, and desynthesize attacks) attacks. 
However, the success of Oracle-based and Oracle-less attacks is not always guaranteed. 
Further, the foundry can deploy hardware Trojan for extracting the locking key. 
Fig.~\ref{fig:foundry_model} summarized the assets and capabilities of a foundry and corresponding attack methodologies of an untrusted foundry.  

\subsubsection{SoC Designer}
An SoC designer has access to the soft/hard IP core, knowledge about the functionality of each IP, and unlocked functional obfuscated chip. 
Besides, the design undergoes extensive functional analysis for bug detection. 
Furthermore, a rogue designer may have access to DFT structures like the scan chain. 
The integrator also has the knowledge of synthesis tools. The capability of the SoC integrator can also include state-of-the-art FA tools and netlist reverse engineering software. 


The primary intention of a malevolent SoC designer for attacking an obfuscated IP is IP piracy/theft. 
A rogue design house may report a less number of chips to the IP owner or clone the IP for selling it to other OCM. 
Hence, 3PIP vendors always have trust issues with the SoC integrator. 

With the aforementioned assets, performing an hardware Trojan insertion, Oracle-guided and Oracle-less attacks on the chip is more convenient for an SoC designer. 
Aside from black-box analysis, a rough SoC integrator with access to reverse engineering and FA tools can also deploy physical attacks like optical probing.

\subsubsection{3rd Party Design Service Provider}
As described in Sec. \ref{sec:supply_chain_1}, in the current SoC design flow, the 3rd party design service provider has complete access to the gate-level netlist as well as the scan chain implemented in the device. 
Besides, the 3rd party design service provider can also gain access to an activated chip.
Their capability may also include netlist reverse engineering and access to FA lab. 
The goal for attacking the hardware obfuscation for a 3rd party service provider is hardware Trojan insertion, IP piracy, and IP overuse. 
Due to access to similar assets like SoC designer, exploiting Trojan,    Oracle-guide and Oracle-less attacks is more convenient for 3rd party service provider. 
Furthermore, they can apply tools used for FA to extract the key value for logic obfuscation or FSM locking. 
The selection of attack methodologies depending on assets and capability for SoC designer and 3rd party design service provider is depicted in Fig.~\ref{fig:socdesigner_model}. 

\subsubsection{End User}
The threat of end user is the most overlooked concern in hardware obfuscation. 
The reason behind such an assumption is a common perception that full-blown reverse engineering is an expensive and expertise oriented process. 
In recent years, advancements in the reverse engineering process should compel the research community to revisit the threat of IP piracy by end users. 
An end user only has access to the unlocked chip and documentation related to that design. 
However, she can delayer each layer, image those layers with SEM and extract the gate-level netlist using reverse engineering software like Pix2Net or Chipwork. 
Even without having access to FA tools and reverse engineering capabilities, an end user without reverse engineering capability can still exploit the design vulnerabilities for extracting key value of FSM or logic locked circuitry using side-channel analysis and probing methods. 

The potential adversaries for hardware obfuscation, their access to assets, their capabilities, and possible attack methods are summarized in Fig.~ \ref{fig:threat_model}. The possible access to capabilities and possible attack methods in Fig.~\ref{fig:threat_model} are ranked from the easiest to hardest.

\section{Architecture for Defense-in-depth} \label{sec:model}
The objective of logic obfuscation is to protect the functionality and design implementation of the chip.
Therefore, the unlocking key in structural obfuscation is considered as the center of attacker interest. 
A designer can select number of defense layers for protecting the locking key, depending on the security budget, design constraints, i.e., area, power, timing constraints. 
Besides, the threats and vulnerabilities of the core components and supply chain also define the security layers implemented in the locked chip.
Furthermore, a security designer must consider the fact that, once an adversary finds the defense mechanism implemented in the device, the attacker has an unlimited amount of attempts to find a hole in the security layer. 
Hence, failure in one defense layer may impact and even sacrifice the integrity of other defense layers. 
For example, success in structural reverse engineering allows a hacker to identify suitable point of interest (PoI) for probing and even expose the defense mechanism implemented against the electrical or optical probing attacks.
In our model we have considered six layers of security for securing the key in hardware obfuscation as shown in Fig.~\ref{fig:defese_in_depth}.
 

\paragraph{Layer --1: Hardware Assurance} The security of the logic locking is established on the assumption that the hardware is secured. 
Any malicious modification detected in the design violates the assumption for root of trust, as well as impose dire threat towards the assets protected in the device.
Hardware Trojan can also weaken the security mechanism implemented in the chip. 
Therefore, establishing trust and assurance on the device should be the first step towards developing the defese-in-depth for logic locking.

\paragraph{Layer -- 2: Defense against Reverse Engineering} Defense against reverse engineering, both structural and information, is considered as the first line of defense for the obfuscated chip.
Attacking an obfuscated chip starts with breaking into the layout obfuscation techniques, learning the implementation of the design and detecting the point of interest for extracting the assets form the device.
Although the cost, time, and expertise are always considered as a challenge for reverse engineering; once the completed reverse engineering attempt exposes valuable information to the adversary. 
An attacker can use that information for completing other attack methods like optical and electrical probing
Hence,protection against structural reverse engineering, increases the complexity of probing, and Oracle-guided attacks. 
Again, from the vulnerability analysis of key storage element shown in Fig.~\ref{fig:framework}, it is also evident, extracting the key value through the information reverse engineering can be a straight forward task for breaking into the logic locking.


.  

\paragraph{Layer -- 3: Defense against Contactless Probing} Once, the adversary knows the location of the key-delivery unit and key-gates from layout reverse engineering, they can raid the key-delivery unit and interconnect layers using contactless method like optical probing from the backside of the chip (see Fig.~\ref{fig:framework}). 
In \cite{cryptoeprint:2019:719}, authors showed the location of key-delivery unit can also be extracted through partial reverse engineering. Due to non/semi-invasive nature of the optical probing, cost and time required for key extraction is much lower than contact-based electrical probing attack. 
The FA tools required for such analysis (laser microscope) can be rented for a few hundred dollars per hours. 
Nonetheless, a modern chip does not have any protection mechanism for the backside of the substrate. 
Therefore, protection against contactless probing has been placed in the second layer in defense-in-depth model. 

\paragraph{Layer -- 4: Defense against Contact-based Probing} Extracting key value from interconnects and key-delivery unit using FIB and electrical probing analysis involves invasive analysis. 
Similar to FA tools used for contactless probing; the tools required for contact-based probing can be rented almost at the same rate. 
However, due to the invasive nature of the attack, the time, cost and expertise required for electrical probing is considered higher than optical probing. 
Although several defence mechanisms have already been proposed, with access to right equipment an adversary can still bypass that defense mechanism. 
Hence, third layer in defense-in-depth should protect the chip assets from FIB/electrical probing (see Fig. \ref{fig:defese_in_depth}). 

\paragraph{Layer -- 5: Defense for Design-for-Test} Literature showed that access to scan chain makes logic obfuscation vulnerable to several scan-based, Oracle-guided and Oracle-less attacks (See Fig.~\ref{fig:framework}).
However, the access is constrained to certain stakeholders which have been discussed in Sect.~\ref{sec:supply_chain}, hence, the protection of the scan chain is placed as the fourth layer in defense-in-depth. 

\paragraph{Layer -- 6: Defense for Logic Obfuscation Techniques} Lastly, logic obfuscation protects the functionality of the design. 
Attacking logic locking techniques requires reverse engineered gate-level netlist, i.e., success in breaking the first line of defense in the obfuscated chip. 
Similar to scan chain attacks, logic locking can also be exploited using Oracle-guided or Oracle-less attack methods to learn the key value (see Fig.~\ref{fig:framework}). 
As the presence of sequential logic poses difficulty against Oracle-guided attack, the defense for logic obfuscation  is placed in the fifth layer of the defense-in-depth. 
\section{Security Measures for Defense-in-depth} \label{sec:countermeasure}

In this section, we will discuss the security measures and future directions for developing defense-in-depth countermeasures for hardware obfuscation for major elements in chip design, i.e, the key-storage interconnect, key-delivery unit, DFT, and obfuscation techniques. 
\subsection{Hardware Assurance}
Detecting malicious modification in the design is the main objective of hardware assurance layer in multi-layer defense approach. 
Several hardware Trojan detection techniques, e.g., run-time monitoring, test based approach, side-channel fingerprinting, have already been proposed to ensure the root of trust for the device~\cite{trojantehranipoor2010survey_of_ht}.
However, none have proved to equally effective or limited due to golden chip requirement, time and memory consumption, process variation, subject matter expert involvement, etc.

Reverse engineering can be an effective means for verifying the the trust and assurance of a chip fabricated in an untrusted foundry. 
However, the application of revers engineering is limited by the lack of automation and invasive nature of the method. 
The time and resources required for Trojan detection can be further reduced by applying computer vision and machine learning approach.
In~\cite{vashistha2018trojan} authors suggested that, A fast SEM image collected from the backside thinned IC can be compared with the golden layout available to the designer for detecting potential malicious circuitry.
In this case, Supervised machine learning and image processing is used to compare the DUA and golden layout. 
A security designer can also insert golden gate circuits (GCC) in the unused space of the design and use the GCC to improve the accuracy of machine learning classifier for detecting the any suspicious modification in the SoC~\cite{shi2019golden}. 
The aforementioned techniques for hardware assurance can prevent the asset leakage like locking key. 
However, meeting the aggressive time-to-market requirement can still be a challenge for the OCM. 
\subsection{Defense against Reverse Engineering}
\label{sec:memory}
\subsubsection{Security Countermeasures}
The defense against the reverse engineering evolves around two core components in the obfuscated IC -- key-storage and obfuscated hardware. 
Here, the protection mechanisms of those core components are reviewed. 
\paragraph{Protecting Key-storage from Reverse Engineering} Developing a secured key-storage device is still a topic for extensive research.
Over the past decades, researchers have proposed several methods to protect the NVM memory from reverse engineering.
Memory encryption can be a solution against key-storage reverse engineering. 
In fact, memory encryption techniques may be the topic of most research activity aimed for protecting the data stored in main memory. 
Encryption algorithms allow strong diffusion characteristics that ensure a single bit change in the plaintext results in several bit changes in the cipertext.
Therefore, an attacker can retain the key persists in the NVM, but in an unintelligible form. 
Although such encryption prevents reverse engineering, the designer should also consider the twofold of vulnerabilities introduced by the memory encryption. 
The decryption method would increase the read latency for key-storage which will adversely affect the performance of the chip. 
Again, the decryption key is also available in the chip which introduces the vulnerability with side-channel attack and introduces vulnerability for the key-delivery unit. 

Anti-fuse technology is a promising solution as secure key-storage due to difficulty in localizing and reading the stored values in anti-fuse. 
This is a mature technology used in FPGAs and PLAs.
Memory cell with different threshold voltage is also proposed as a possible key-storage cell.
Using controlled process variations like dopant value, the threshold voltage of manufactured transistors can be varied from nominal values.
Later the variation in threshold voltage is used to define the output from a logic cell~\cite{keshavarz2017threshold}. 
Nonetheless, before using this method potential vulnerabilities against  SEM, PVC, and other charge probing techniques should be addressed to block the reverse engineering of NVM. 

Emerging NVM memory technologies can be considered as possible alternatives of the existing key-storage like Flash, EEPROM. 
Emerging memories -- resistive random access memory (RRAM), spin-transfer torque magnetic random-access memory (STT-RAM), phase change memory (PCM) do not use the charge as storage media. 
For example, RRAM typically operates by electrical switching between different resistance states by applying high voltage, observed in several metal oxides~\cite{xie2016logic}. 
Applying high voltage across the metal plates switches resistance states of the device. 
The high resistance state is considered as bit ’1’ and the low resistance state is considered as bit ’0’. 
As there is no visual difference between the bit '1' and '0', it is difficult to extract the stored value from memory. 
Therefore, the aforementioned memory technologies are protected against the conventional charge probing techniques like SKPM, SCM, PVC.  
However, the susceptibility of the aforementioned memories against the side-channel analysis, or other types of probing (for example, EBIC/EBAC), or microscopy (for example, spin-SEM) should be evaluated.       

\paragraph{Protecting Obfuscated Hardware from Structural Reverse Engineering}
Several countermeasures have been proposed to protect IC camouflaging against SAT, brute-force, and sensitization attacks. 
In~\cite{yasin2016camoperturb}, authors have proposed to perturb the functionality of the given design minimally by adding or removing one minterm. 
A camouflaged block, CamoFix, built up using camouflaged inverter/buffer cells, is used to restore the perturbed minterm in the functionality of the design. 
However, these techniques are vulnerable to removal attacks~\cite{yasin2017removal}. 
Researchers have also proposed to use layout-inclusive interconnect locking scheme based on cross-bars of metal-to-metal programmable -via devices.  
Logic locking scheme using antifuse to connect two adjacent metal layer proposed in~\cite{shamsi2018cross}, incorporated dummy vias and filler cells to eliminate the requirement for secure key storage. 

Another solution for camouflaging the gate is to use different threshold voltage defined (TVD) logic gates~\cite{erbagci2016secure}. 
The TVD logic gates are implemented with different threshold voltage transistors by varying the doping implant in the transistor. The gates have an identical layout, however, the threshold voltage defines the functionality of each gate. 
Covert gate is another variant for camouflaging cells~\cite{shakya2019covert}.  
Variation in doping concentration and dummy contacts are used to develop covert gates that are indistinguishable from regular logic gates in a design.
Further, the gates shows higher resistance to SAT attack unless the location of the covert gate is identified. 


\subsubsection{Future Directions}
Confidentiality of unlocking key value significantly affects the security of structural obfuscated IP/chip. 
Therefore, a key-storage should have the following three properties: 
\begin{enumerate} [label=(\alph*)]
\item The key-storage must be read-proof, i.e., malicious entity cannot reverse engineer or extract the key value form the storage.
\item The key-storage must be tamper-evident, i.e, it can detect tampering attempts and zerozied the content irrespective power status of the chip.
\item The key-storage introduces lower area or power overhead to the chip to be an effective solution for chip. 
\end{enumerate}
Developing a framework for accessing the attack resiliency to different memory technology can be a contribution for the research community. 
In recent years, several emerging memory technologies have been proposed as a possible selection for secured key-storage. 
The performance of those memory technologies against known attack methods -- invasive, non-invasive or semi-invasive methods is yet to be evaluated. 
Again, protecting the backside of NVM from unauthorized access can contribute to protecting the hardware obfuscation. 
Developing active or tamper-evident shield to protect the memory can also be a significant advancement towards securing the key-storage. Developing a light key encryption algorithm can thwart exposing the key.
Furthermore, several other questions needed to be answered like how to overcome the bottleneck due to read latency for key-storage, erasing the residual data, and masking the location of OTP from advanced imaging tools. 

The challenge of developing physical layout obfuscation technique is area, power, and delay overhead incorporated with the camouflage cells.
Besides, developing threshold dependent camouflage cells involves dopant variation which can be identified from SEM imaging of the die at different beam voltages. 
Programming the TVD logic gates at the post-manufacturing stage has also been proposed as a possible camouflage technique~\cite{akkaya2018secure}. 
The scalability of the TVD logic gates is always a concern for the semiconductor industry.

\subsection{Defense against Contactless Probing}
 \begin{figure}[!t]
\centering
\includegraphics[width=0.45\textwidth]{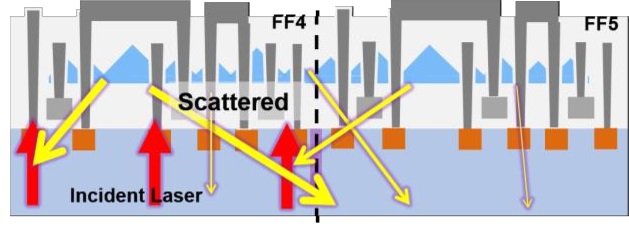}
\caption {Scattered reflection of incident laser beam in a nanopyramid implemented device~\cite{shen2018nanopyramid}.}
 \label{fig:nanopyramid}
\end{figure}

Security against optical analysis mostly concerns protecting the backside of the chip. 
Backside protection of the chip has received more attention recently from the security research community. 
The possible countermeasure for the backside of a chip can be divided into two levels -- device and circuit level.  

A security designer can add a backside polishing detector to monitor the thickness of the bulk silicon existing below the transistor. It has already been proposed as a countermeasure against mechanical polishing~\cite{manich2015backside}.
Adding an active opaque layer can be another countermeasure against optical probing.
Implementing an active monitoring scheme is required to detect the removal of such opaque layer by an  adversary~\cite{amini2018assessment}. 
Since the optical beam stimulates the silicon active regions thermally, conventional photosensors fail to trigger during optical probing. 
On the other hand, the thermal simulation introduces temperature and current variations in the circuit, which can influence circuits, such as ring-oscillators (ROs)~\cite{tajik2017pufmon}. 
In this case, the implementation of ROs as a probing protection scheme can be used to generate an antitamper reaction in the chip to protect the locking keys. 
In~\cite{shen2018nanopyramid}, nanopyramid structures are implemented in selective areas inside the chip to mitigate optical probing attacks by scattering the reflected laser beam, and consequently, scrambling the measurements of the register contents. 
Another proposed countermeasure is implementing a sandwiched metal shield between two polymers, opaque to NIR, at the back of the chip.
As the layer can be removed using acid etching or polishing the chip, associating the stability of the bulk silicon to that sandwiched layer is required to prevent the adversary from taking off~\cite{borel2018novel}.  

A circuit-level solution can be widely accepted for the semiconductor industry. 
As the logic locking key is static and embedded in the device memory, it can
be probed by the aforementioned attacks. As a solution, the IP owner can use dummy active registers connected to functional gates to disguise the key-registers and eventually hide the key-gates. 
Although, the circuit level countermeasures might be known to a
malicious foundry, and they can be easily deactivated, these countermeasures can be considered more secure against end users.

\begin{figure}[b]
\centering
\includegraphics[width=0.4\textwidth]{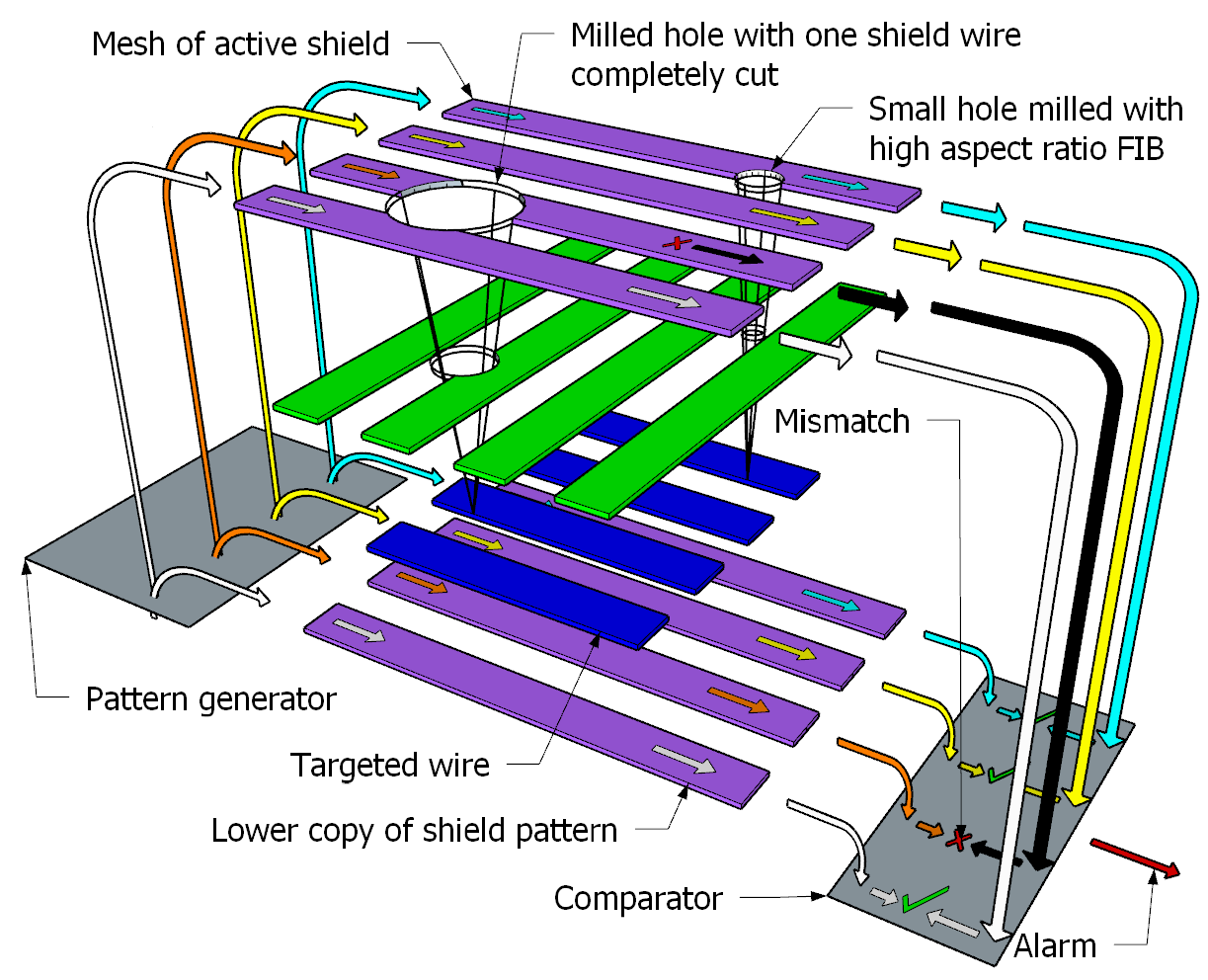}
\caption {Working principle of active shield and bypass attack on active shield~\cite{Wang2019}.}
\label{fig:active_shield}
\end{figure}

\subsection{Defense against Contact-based Probing Attacks}
\label{sec:interconnects}



Active shield, which is also called digital shield, is the most common countermeasure against front-side probing attack~\cite{randomshield2012,CryptoShield2017}. In active shield technology, a signal carrying shield is placed on the top layer(s) of the chip to detect whether one of the shield wires is cut or not as shown in Fig.~\ref{fig:active_shield}.
A pattern generator is required for an active shield to generate flipping patterns to be transmitted on the shield. 
Then, a comparator at the end of the shield compares the received pattern from upper shield wires and another shield pattern copy from lower layers.
If there is a mismatch detected at the comparator, which means at least one of the shield wires are cut during the attack, an alarm will be triggered, e.g. erasing all sensitive data stored in memory.
The generated pattern should not be predicted or controlled by the attacker since if the shield patterns are compromised, then the attacker can synchronize the pattern at the end of the shield using fault injection techniques. Therefore, the shield wires before the fault injection sites are free to cut, which results in that the integrity checking function of the active shield is totally disabled.
Although the active shield is very popular, its large design overhead and vulnerability to advanced FIB system limit the wide application of it.
First, a naive active shield on the top layer is very vulnerable to reroute and bypass attack with advanced FIB system as illustrated in the previous subsection.
Then, the active shield typically occupies one entire routing layer which is prohibitively expensive to designs with tight cost margin and technologies with few routing layers.
Further, the requirements for a non-predictable and non-controllable pattern generator determines that it is not a simple and small component, e.g. a cipher-based pattern generator with finite state machine (FSM) as its input, which introduces large area and power overhead to the design, especially when the design itself is relatively small, such as an AES or DES encryption core.
In addition, the attacker can also utilize FIB's circuit editing capability to manipulate the control circuit and payload of active shield to disable it.

\begin{figure}[!t]
\centering
\includegraphics[width=0.45\textwidth]{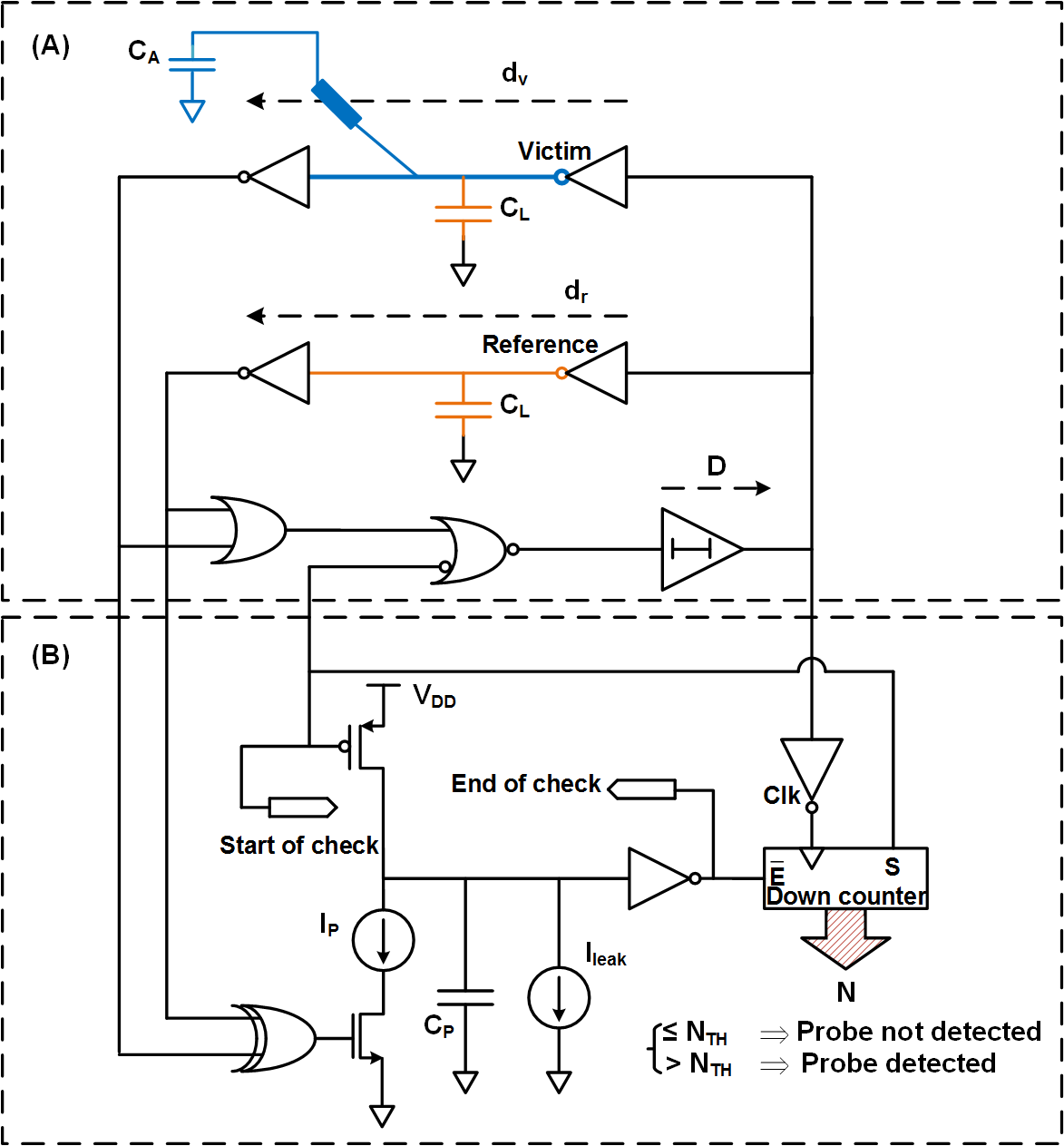}
\caption {Probe Attempt Detector (PAD)~\cite{Wang2019}.}
\label{fig:PAD}
\end{figure}

Analog shield and sensors are alternative approaches to active shield~\cite{PAD2012, PAD2018}. 
Unlike active shield which detects the attack by comparing digital patterns, analog shield and sensors utilize analog features, e.g. capacitance or RC delay, at specific chip locations to detect the attack.
One example is the Probe Attempt Detector (PAD) \cite{PAD2012} as shown in Fig.~\ref{fig:PAD}.
It detects the attack by measuring the additional capacitance introduced by the probe on selected sensitive wires.
Compared to active shield which is covering a large chip area, the PAD approach is wire-oriented which is difficult to be applied to a large group of sensitive wires.
Therefore, if only a few wires are identified as security-critical wires and need to be protected, PAD is a good option with small overhead.
Another example is charge sensor~\cite{ChargeSensor2012} which detects the attack by sensing charges during the FIB navigation process before the exact milling.
An extremely sensitive local charge sensor is placed close to the chip surface, which could capture the charge changes and store the state for later read-out.
However, the charge sensor accuracy is limited by the environmental noise and other power, voltage, and temperature (PVT) variations.
Further, the charge sensor is not working in real-time, which leaves opportunity for attackers to neutralize the charge before the read-out of the stored state in charge sensor.
In addition, one common and main limitation for all analog-based countermeasures, which typically requires a threshold value to trigger an alert, is that they are less reliable against process variation. 
It is very difficult to distinguish between an attack and a reasonable process variation when the attacker's footprint is getting smaller and smaller with advanced equipment.

\begin{figure}[!t]
\centering
\includegraphics[width=0.45\textwidth]{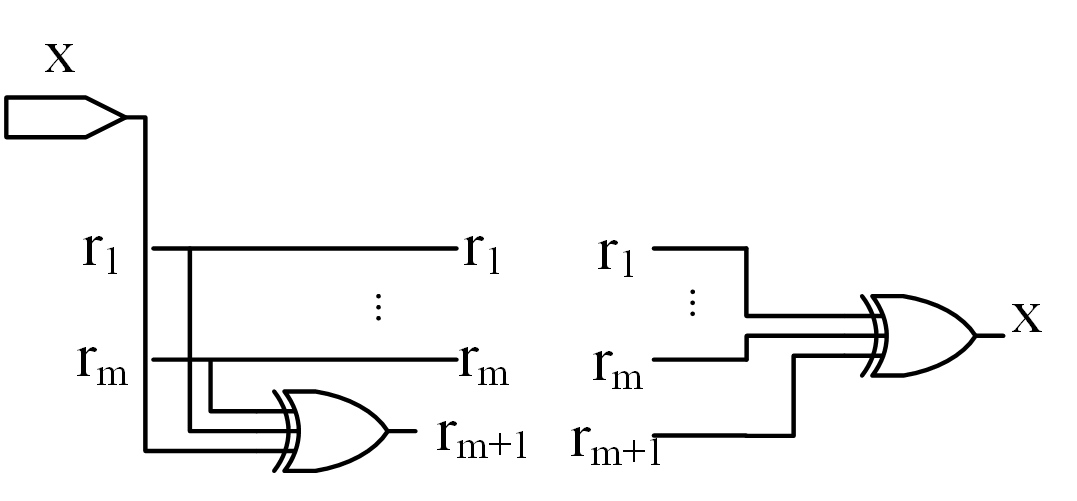}
\caption {Input encoder (left) and output decoder (right) for
masking in t-private circuit~\cite{wang2016probing}.}
\label{fig:t_private}
\end{figure}

Different from active shield and analog sensor which are designed to detect the probing attack, $t$-private circuit \cite{t_private} is proposed to deter the attack by exhausting the number of simultaneous probes in a probe station system which typically has 4-8 concurrent probes, so that the attacker doesn't have enough concurrent probes to extract one bit of information.
Fig.~\ref{fig:t_private} shows the diagram of $t$-private circuit which transforms the one bit signal $X$ to m+1 bit signals ($r_1, r_2, ... r_{m+1}$), so that at least m+1 probes are required within one clock cycle to extract one bit signal $X$.
When m+1 exceeds the number of probes that the system provides, it would be very difficult for attackers to extract sensitive information through the probing attack.
The main issue with $t$-private circuit is that the area overhead involved for transforming all signals in a chip is prohibitively expensive ($O(t^2)$). The scheme also requires a random bitstream generated at every clock cycle for the signal transformation. 

To sum up existing countermeasures, we can find that every single solution is not efficient enough to resist probing attack and has its limitations.
So, we need a holistic and efficient solution against probing attack urgently because attacker's capability is always improving with advanced techniques.
We believe that the following directions and suggestions are worth putting more effort to improve current countermeasures against the probing attack that can extract sensitive information from chip interconnects.

Security designers should keep in mind that a successful probing attack consists of many steps, such as navigation, milling, depositing, data extraction, etc.
Do not only focus on the milling step, like most shield-based countermeasures.
If we can efficiently detect or deter two or more necessary steps in an attack, we could improve our protection performance and confidence to a great extent.

With the rapid improvement of the attacker's capability, especially for those attackers equipped with advanced FIB, FIB's capabilities, features, and limitations should be well modeled and considered in the countermeasure development.
For example, FIB's aspect ratio, which is the ratio between depth and diameter, should be considered in a shield-based countermeasure. It is  because of the fact that the width and space of the shield wires and the depth difference between shield layer and probing target layer could determine if the shield is useless for a FIB system whose aspect ratio is larger than a specific value.

Almost all existing countermeasures have scalability issue with large overheads in chip area and layers and performance degradation, which is not acceptable for most high-end chips, e.g., CPU, that have a very limited budget for security.
Therefore, a highly efficient solution is needed to protect those most sensitive nets in the design with minimal overhead.

In addition, there is no effective countermeasure against back-side probing occurring from the substrate of the chip which might be more threatening than front-side probing because it is much easier to get access to the transistors and sensitive nets on lower layers from the backside.

\subsection{Defense for Design-for-Testability}
A number of countermeasures have been proposed in the literature so far to thwart scan-based oracle-less and oracle-guided attacks. A brief discussion of existing techniques is given below.

Dynamically Obfuscated Scan (DOS): The authors in \cite{wang2018secure} proposed a design and test methodology against scan-based attacks throughout the supply chain, which includes a dynamically obfuscated scan for protecting IP/ICs. By perturbing test patterns/responses, and protecting the obfuscation key, the proposed architecture is proven to be robust against existing noninvasive scan-based attacks and can protect all scan data from attackers in the foundry, assembly, and system development, without compromising the testability. The key difference of this technique from other countermeasures is, rather than using a static obfuscation key, authors have proposed a dynamic obfuscation approach where the obfuscation key changes periodically based on the given permutation rate and initial seed of the LFSR \cite{wang2018secure}, making the overall design resistant to scan-based side-channel attacks.

Low-Cost Secure Scan (LCSS): In \cite{hely2005test}, authors have presented the low-cost secure scan (LCSS) solution. LCSS is implemented by inserting dummy flip-flops into the scan chains; it inserts the key into the test patterns with respect to the position of the dummy flip-flops in the chains. By doing so, it verifies that all vectors scanned-in comes from an authorized user, and the correct response can be safely scanned-out after functional mode operation. If the correct key is not integrated into the vector, an unpredictable response is scanned-out, making analysis very difficult for an attacker. By using an unpredictable response, attackers would not be able to immediately realize that their intrusion has been detected, as could be discerned if the CUT were to immediately reset \cite{hely2005test}.

Lock and Key: Lock \& Key solution was developed to neutralize the potential for scan-based side-channel attacks \cite{lee2005securing}. The Lock \& Key technique provides a flexible security strategy to modern designs, without significant changes to scan structure used in practice. Using this technique, the scan chains in a SoC are divided into smaller sub-chains. With the inclusion of a test security controller, the values of access to sub-chains are randomized when being accessed by an unauthorized user. Random access reduces repeatability and predictability, making reverse engineering more difficult. Without proper authorization, an attacker would need to unveil several layers of security before gaining proper access to the scan chain in order to exploit it.

Obfuscated Scan: Secure scan architecture using test key randomization (SSTKR) was developed to address security and testability issues \cite{razzaq2011sstkr}. Specifically, SSTKR is a key-based technique to prevent an attacker from illegally obtaining critical information while using scan infrastructure. The authentication keys are generated through linear feedback shift register and inserted into test vectors. Furthermore, test keys are embedded into test vectors in two different ways: with dummy flip-flops and without dummy flip-flops. In the first case, dummy flip-flops holding the key are inserted into the scan chain to randomize scan outputs. It should be noted that all dummy flip-flops should not be connected to the combinational logic. 
In the second case, authentication keys are inserted into the positions of don’t-care bits, generated by ATPG to reduce area overhead and test time.

Scan Encryption: A countermeasure against scan-based side channel attacks could be done through the encryption of the scan chain content \cite{da2017scan}. These attacks use an efficient and secure block cipher placed at each scan port to decrypt/encrypt scan patterns/responses at each scan input/output, respectively.

Scan-chain Reordering: A secure scan tree architecture is developed to protect cryptosystems against scan-based attacks \cite{sengar2007secured}. This architecture offers low area overhead compared with the traditional scan tree architecture followed by a compactor, locking, and test access port (TAP) architecture. In contrast to the normal scan tree architecture, this architecture is based on the flipped scan tree (F-scan tree). To be exact, they adopt special flip-flops (that is, flipped FFs) in which inverter gates are added at the scan-in pin of scan flip-flop. The flipped scan tree architecture is built through normal SDFFs and flipped FFs. Since the attacker cannot identify the position of inverters, he/she is neither able to control the inputs, nor observe the outputs of the flip-flops.
\subsection{Defense for Logic Obfuscation Techniques}

Although logic obfuscation can be an effective mechanism for establishing trust in the hardware design flow, it has not seen a widespread application in the semiconductor industry due to its lack of attack resiliency and formal notion of security. 
For example, most logic obfuscation techniques are vulnerable to SAT-based attacks \cite{subramanyan2015evaluating}.


To resist SAT attack, several SAT-resistant logic obfuscation techniques~\cite{yasin2016sarlock,xie2018anti,yasin2017provably} can be implemented in the IP design. 
These SAT-resistant logic locking techniques that increase SAT attack complexity by increasing the number of required distinguishing input patterns (DIPs) \cite{yasin2016sarlock,xie2018anti} or by striping some of the functionality of the logic locked design \cite{yasin2017provably} and hiding it in the form of a secret key, possess their own critical vulnerabilities. 
For example, researchers have proposed Bypass attack
~\cite{xu2017novel}, SPS attack~\cite{yasin2017removal}, App-SAT attack~\cite{shamsi2017appsat}, and FALL attack~\cite{subramanyan1functional} that can easily circumvent the effect of the SAT-resistant locking schemes. 
Further, these SAT-resistant schemes are known to possess low corruptibility, and thus do not provide the desired functional obfuscation. 
Hence, there remains a need for developing SAT-resistant logic obfuscation infrastructure. 
Since SAT attack relies on access to scan chain, effectively obfuscating/locking the scan chain to scramble scan-in and scan-out should help resist such attacks. 
A recently proposed scan architecture~ \cite{wang2018secure} resists bypass, reset, flushing and other scan-based attacks by dynamically obfuscating scan chain where scan chain obfuscation key changes periodically. 
This idea of dynamically changing obfuscation key can also be utilized to resist SAT attack. 
SAT attack requires access to unlocked IC (oracle), locked netlist and a number of iterations to rule out incorrect keys. 
If the obfuscation key can be changed each time before SAT attack succeeds, then attack complexity would be drastically increased.

Developing key-gate insertion algorithm to improve the output corruptibility for wrong keys as well as thwart attacks similar to key-sensitization~\cite{rajendran2012security} can improve the defense mechanism of the logic locking. A fault analysis (FA) based key-gate insertion algorithm has already been proposed~\cite{rajendran2015fault}. Increasing the dependency among the keys in the key-gate placement is also explored in strong logic locking (SLL) algorithm~\cite{yasin2016improving}. However, all these algorithms are vulnerable to key-sensitization, or logic cone based~\cite{lee2015improving} attacks. Moreover, key-gate insertion algorithm can only be successful to protect the key value, if the chip is protected from reverse engineering and probing (both contactless and contact-based methods) attacks.

   \section{Research Opportunities} \label{sec:future_research}
There is no single silver bullet solution for addressing all the vulnerabilities in logic locking. Thus, multi-layer protection for the logic obfuscation is necessary to prevent an attacker from stealing the design secret. Although in this paper a possible framework for planning and selection of the defense layers has been laid, several other questions are yet to answer. 

Critical challenges in developing a multi-layer defense mechanism is to select the appropriate countermeasure that can address the corresponding threat in a comprehensive manner. The security designer has to decide the countermeasures to implement for each defense layer. Identifying the security metric and security rule check for defense layers can address the issue of countermeasure selection. For example, the designer can enumerate all known alternative safeguarding techniques for contact-based electrical probing technique and estimate the cost and time required for breaking into the defense layer using the metric developed. Furthermore, involving the attacker capability is also necessary for developing a framework for the assessment of security metrics. Therefore, developing a framework to analyze the vulnerabilities and assessing security of the design at all design stages can be a huge contribution in selecting the countermeasures for each defense layer. 

Another factor for defense-in-depth implementation is the allocation of security budget in terms of area, speed, power, and design cost for any specific embedded device. Such analysis enables the integration of the functional and countermeasure design in a holistic fashion. Moreover, the reliability of countermeasure is also dependent on the speed, power, and temperature variation such as sensor-based optical probing detector may not be able to detect low power laser beam if the security constraints are not selected properly. The sensor may ignore the local increase in temperature while optical probing is carried out at low laser power. For most of the time, the attack level and available security budget for a specific product are correlated. A high-end product with high IP value may be confronted with attackers with the most advanced equipment, and thus, may have more budget to adopt more countermeasures in the design. Therefore, when the IP value and the attack threat of a product can be accurately estimated,  the security designers can have more clues to determine which protection technique can be incorporated in the design.

\section{Conclusion} \label{sec:conclusion}

In this paper, we have presented a comprehensive study of different vulnerabilities of the core components, i.e., key-storage element, key-delivery unit, interconnect, DFT and structural obfuscation; in hardware obfuscation. 
Hardware obfuscation is emerging as a promising tool for protecting the IP/chip design and root-of-trust.
Therefore, the dire threat imposed by the vulnerabilities of hardware obfuscation core components in an SoC can not thwart by one-to-one protection scheme. 
Advancement in FA tools and algorithm based attacks do not leave scope to consider any specific protection scheme as the ultimate preserver of the confidentiality and integrity of chip design.
Through using multiple safeguard techniques to protect core components can defend the obscured chip from a variety of attacks. 
Therefore, this paper introduced the idea of a multi-layered defense mechanism which can ensure defense-in-depth for the chip security. 
We have presented the contribution of each stakeholder in the supply chain of the semiconductor device. 
The outline for comprehensive threat model is also presented considering the possible capabilities and assets availability for all possible untrusted entity in the supply chain. 
On the basis of the above analysis we proposed a multilayer defense structure to establish the defense-in-depth in the IC. 
We also discussed about the state-of-the-art defense mechanism for each layer and challenges for paving the path of the secured chip for developing a multi-layer protection scheme. 
Addressing the challenge of incorporating  the    multi-layer defense mechanism can be a significant advancement in the field of logic obfuscation.

\balance
\bibliographystyle{IEEEtran}
\bibliography{main.bib}

\end{document}